\documentclass[sigconf,authorversion=true]{acmart}

\usepackage{graphicx, caption, subcaption}
\usepackage[show]{chato-notes}
\usepackage{amsmath}
\usepackage{epsfig}
\usepackage{algorithm}
\usepackage{algpseudocode}
\usepackage{xpatch}
\usepackage{xspace}
\algrenewcommand\alglinenumber[1]{{\sffamily\footnotesize#1}}

\makeatletter
\xpatchcmd{\algorithmic}{\itemsep\z@}{\itemsep=0.2ex}{}{}
\makeatother

\usepackage{url}
\usepackage{hyperref}
\usepackage{booktabs}
\usepackage{lipsum}
\usepackage[framemethod=TikZ]{mdframed}
\usepackage{multicol}
\usepackage{adjustbox}
\usepackage{tikz}
\usetikzlibrary{bayesnet}
\usepackage{tikz-3dplot}

\hypersetup{
    bookmarks=true,         %
    unicode=false,          %
    pdftoolbar=true,        %
    pdfmenubar=true,        %
    pdffitwindow=false,     %
    pdfstartview={FitH},    %
    pdftitle={My title},    %
    pdfauthor={Author},     %
    pdfsubject={Subject},   %
    pdfcreator={Creator},   %
    pdfproducer={Producer}, %
    pdfnewwindow=true,      %
    colorlinks=true,       %
    linkcolor=black,          %
    citecolor=black,        %
    filecolor=black,      %
    urlcolor=black           %
}

\newcommand{\spara}[1]{\smallskip\noindent{\bf #1}}

\DeclareMathOperator*{\argmin}{arg\,min}

\renewcommand{\algorithmicrequire}{\textbf{Input:}}
\renewcommand{\algorithmicensure}{\textbf{Output:}}

\newcommand{\squishlist}{
 \begin{list}{$\bullet$}
  {  \setlength{\itemsep}{0pt}
     \setlength{\parsep}{3pt}
     \setlength{\topsep}{3pt}
     \setlength{\partopsep}{0pt}
     \setlength{\leftmargin}{2em}
     \setlength{\labelwidth}{1.5em}
     \setlength{\labelsep}{0.5em}
} }
\newcommand{\squishlisttight}{
 \begin{list}{$\bullet$}
  { \setlength{\itemsep}{0pt}
    \setlength{\parsep}{0pt}
    \setlength{\topsep}{0pt}
    \setlength{\partopsep}{0pt}
    \setlength{\leftmargin}{2em}
    \setlength{\labelwidth}{1.5em}
    \setlength{\labelsep}{0.5em}
} }

\newcommand{\squishdesc}{
 \begin{list}{}
  {  \setlength{\itemsep}{0pt}
     \setlength{\parsep}{3pt}
     \setlength{\topsep}{3pt}
     \setlength{\partopsep}{0pt}
     \setlength{\leftmargin}{1em}
     \setlength{\labelwidth}{1.5em}
     \setlength{\labelsep}{0.5em}
} }

\newcommand{\squishend}{
  \end{list}
}

\newcommand{\modelfull}{Multidimensional Ideology-aware Propagation Model\xspace}
\newcommand{\model}{\textsc{MIP}\xspace}

\DeclareMathOperator{\BetaDistribution}{Beta}
\DeclareMathOperator{\Dirichlet}{Dir}
\DeclareMathOperator{\sample}{\textsc{sample}}

\newcommand{\nodes}{\ensuremath{V}}
\newcommand{\items}{\ensuremath{\mathcal{I}}}
\newcommand{\dataset}{\ensuremath{\mathcal{D}}}

\newcommand{\act}[1]{\ensuremath{#1 \in \dataset_i}}

\copyrightyear{2021}
\acmYear{2021}
\setcopyright{acmlicensed}
\acmConference[CIKM '21]{Proceedings of the 30th ACM International Conference on Information and Knowledge Management}{November 1--5, 2021}{Virtual Event, QLD, Australia}
\acmBooktitle{Proceedings of the 30th ACM International Conference on Information and Knowledge Management (CIKM '21), November 1--5, 2021, Virtual Event, QLD, Australia}
\acmPrice{15.00}
\acmDOI{10.1145/3459637.3482444}
\acmISBN{978-1-4503-8446-9/21/11}

\begin{document}
\fancyhead{}

\title{Learning Ideological Embeddings from Information Cascades}

\author{Corrado Monti}
\affiliation{\institution{ISI Foundation, Italy}}
\email{corrado.monti@isi.it}

\author{Giuseppe Manco}
\affiliation{\institution{ICAR-CNR, Italy}}
\email{giuseppe.manco@icar.cnr.it}

\author{Cigdem Aslay}
\affiliation{\institution{Aarhus University, Denmark}}
\email{cigdem@cs.au.dk}

\author{Francesco Bonchi}
\affiliation{\institution{ISI Foundation, Italy \& Eurecat, Spain }}
\email{francesco.bonchi@isi.it}

\begin{abstract}

Modeling information cascades in a social network through the lenses of the ideological leaning of its users
can help understanding phenomena such as misinformation propagation and confirmation bias, and devising techniques for mitigating their toxic effects.

In this paper we propose a stochastic model to learn the ideological leaning of each user in a multidimensional ideological space, by analyzing the way politically salient content propagates.
In particular, our model assumes that information propagates from one user to another if both users are interested in the topic and ideologically aligned with each other.
To infer the parameters of our model, we devise a gradient-based optimization procedure maximizing the likelihood of an observed set of information cascades. Our experiments on real-world political discussions on Twitter and Reddit confirm that our
model is able to learn the political stance of the social media users in a multidimensional ideological space.

\end{abstract}



\maketitle \sloppy

\section{Introduction}
\label{sec:intro}
\enlargethispage{\baselineskip}
The widespread adoption of social-media platforms has altered the landscape of societal debates in unprecedented ways:
an immense amount of content is delivered to social-media users in their timelines, allowing them to quickly access information and participate in political discourse.
However,  with their algorithmically curated and virally propagating content, social-media platforms are suspected of contributing to the polarization of opinions by means of the so-called \emph{``echo chamber''} effect, due to which users tend to interact with like-minded individuals, reinforcing their own ideological viewpoint~\cite{bakshy15exposure, garimella2018quantifying}. Hence, understanding
the interplay between the ideological leanings of social media users and the information they consume and propagate, is of crucial importance towards devising techniques for limiting misinformation, echo chambers, and for designing public information campaigns~\cite{garimella17balancing,aslay2018maximizing}.

Therefore, modeling opinions and their dynamics in social media has attracted considerable interest in the last years.
The bulk of the literature on opinion modeling~\cite{del2017modeling,monti2020learning} and opinion mining~\cite{barbera2015birds,garimella2018quantifying,wong2016quantifying} is however limited to analyze ideological leanings
along a \emph{one-dimensional political spectrum}, i.e.,
the traditional left-right spectrum, or country-specific axis (e.g., United States' Democrat-Republican).
Instead, political scientists have long noted that a single left-right axis assumption is inadequate for describing the existing variation in ideological profiles, and risks to ignore important distinctions between ideological groups~\cite{davis1970expository, eysenck1975structure}. Only recently, opinion dynamics research has started to unveil the complex multidimensional nature of opinion formation~\cite{baumann2020emergence}.

\begin{figure}[t!]
    \centering
    \footnotesize

    \begin{center}
    \resizebox{0.9\linewidth}{!}{
      \begin{tabular}{ cc|c }
      \toprule
      \multicolumn{2}{c}{\textbf{Input} } &  \textbf{Output} \\
      \midrule
      \textbf{Item 1} & \textbf{Item 2} & \textbf{Polarity}\\
      0.9 economy + & 0.9 economy + &      on economy \\
      0.1 minorities & 0.1 minorities &  \\
      \includegraphics[width=0.18\textwidth]{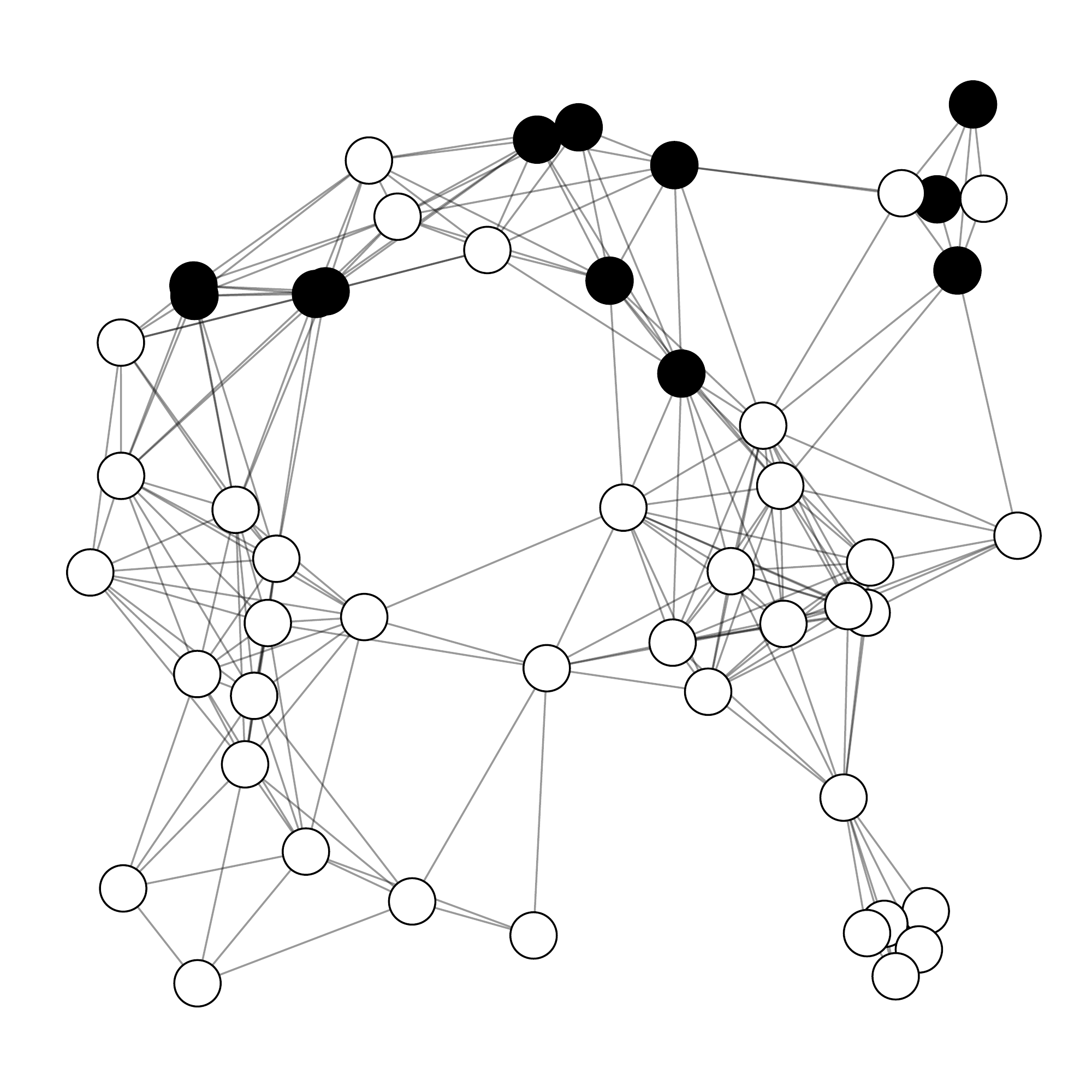} &
      \includegraphics[width=0.18\textwidth]{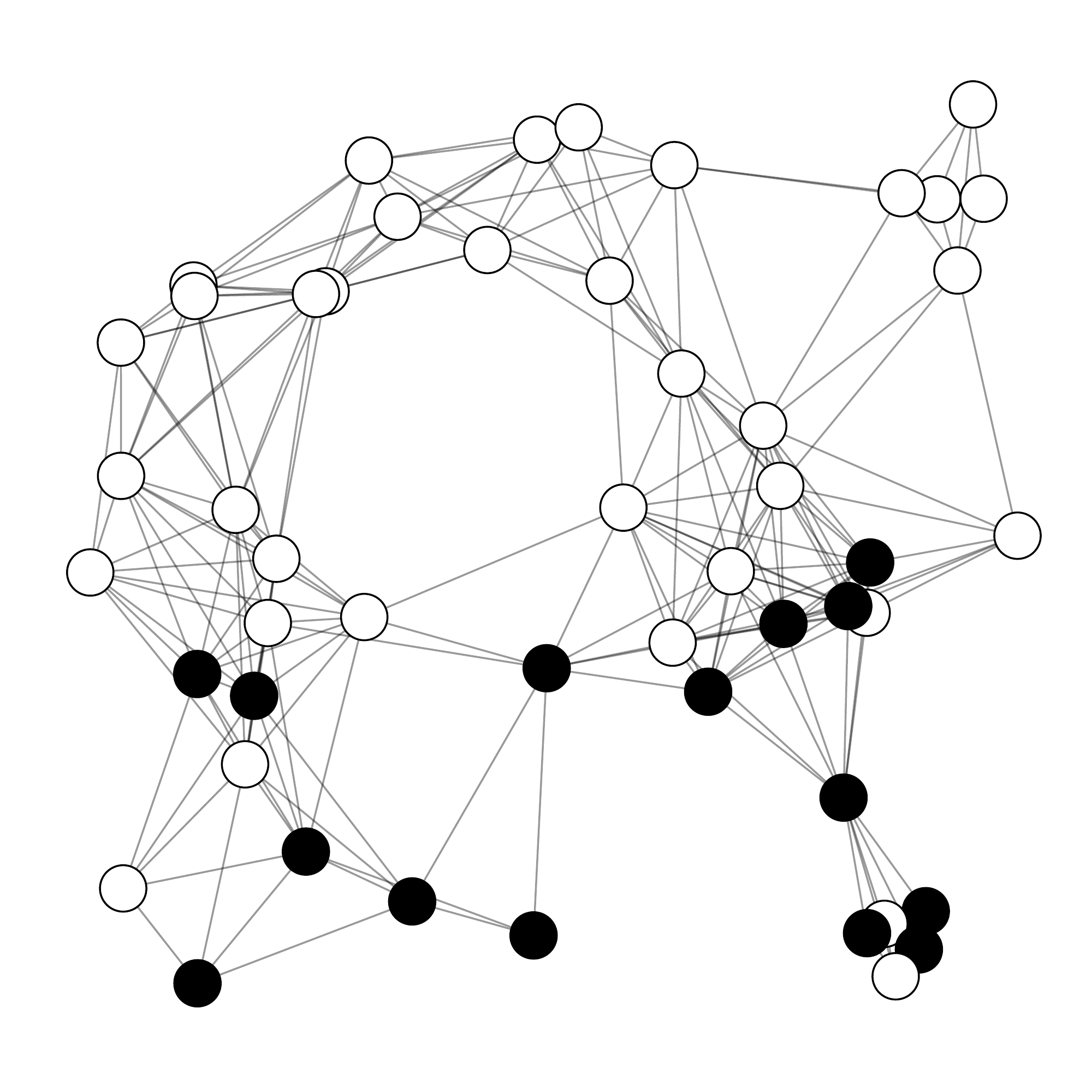} &
      {\includegraphics[width=0.18\textwidth]{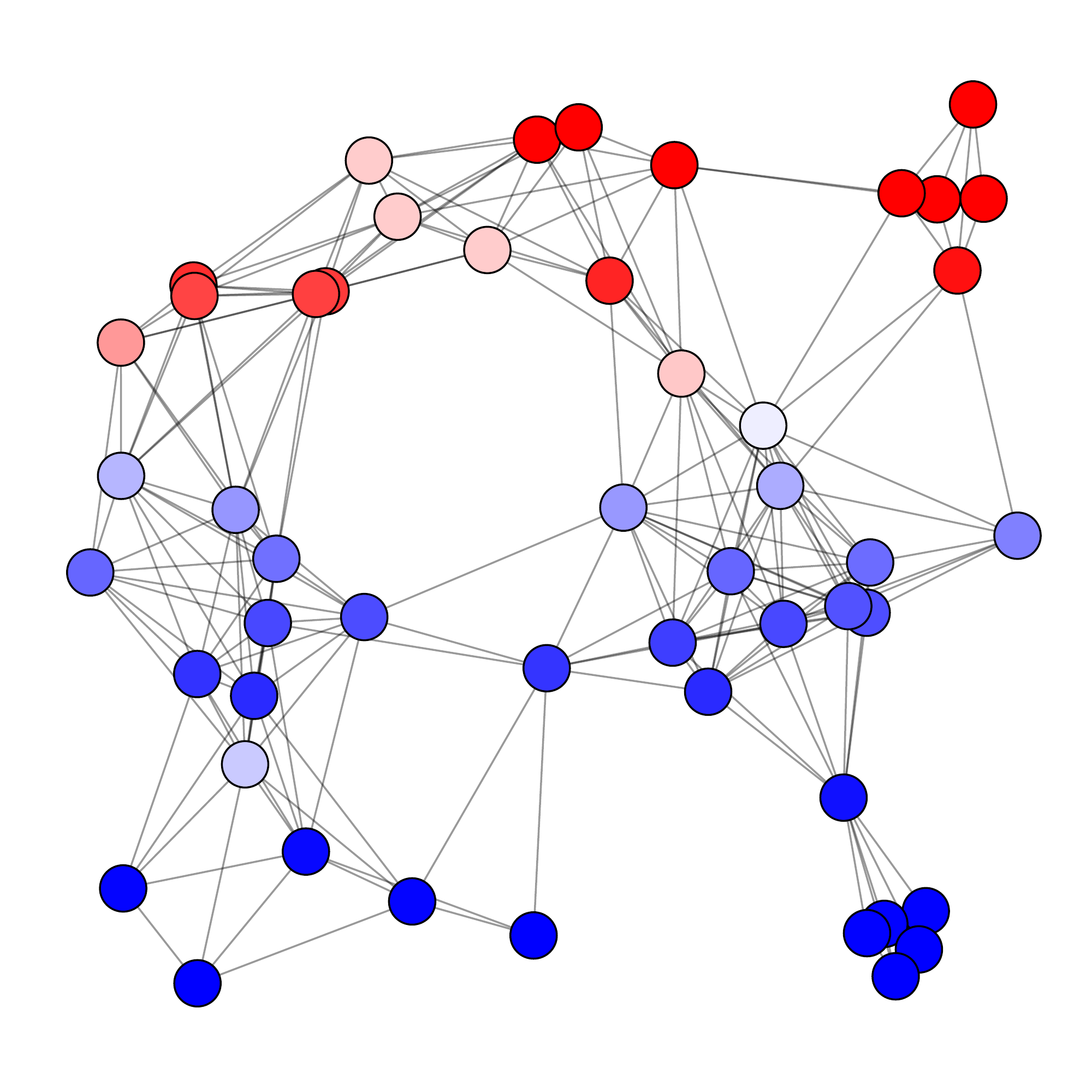}}
      \\
      \textbf{Item 3} & \textbf{Item 4} & \textbf{Polarity}\\
      0.1 economy + & 0.1 economy + &      on minorities \\
      0.9 minorities & 0.9 minorities  \\
      \includegraphics[width=0.18\textwidth]{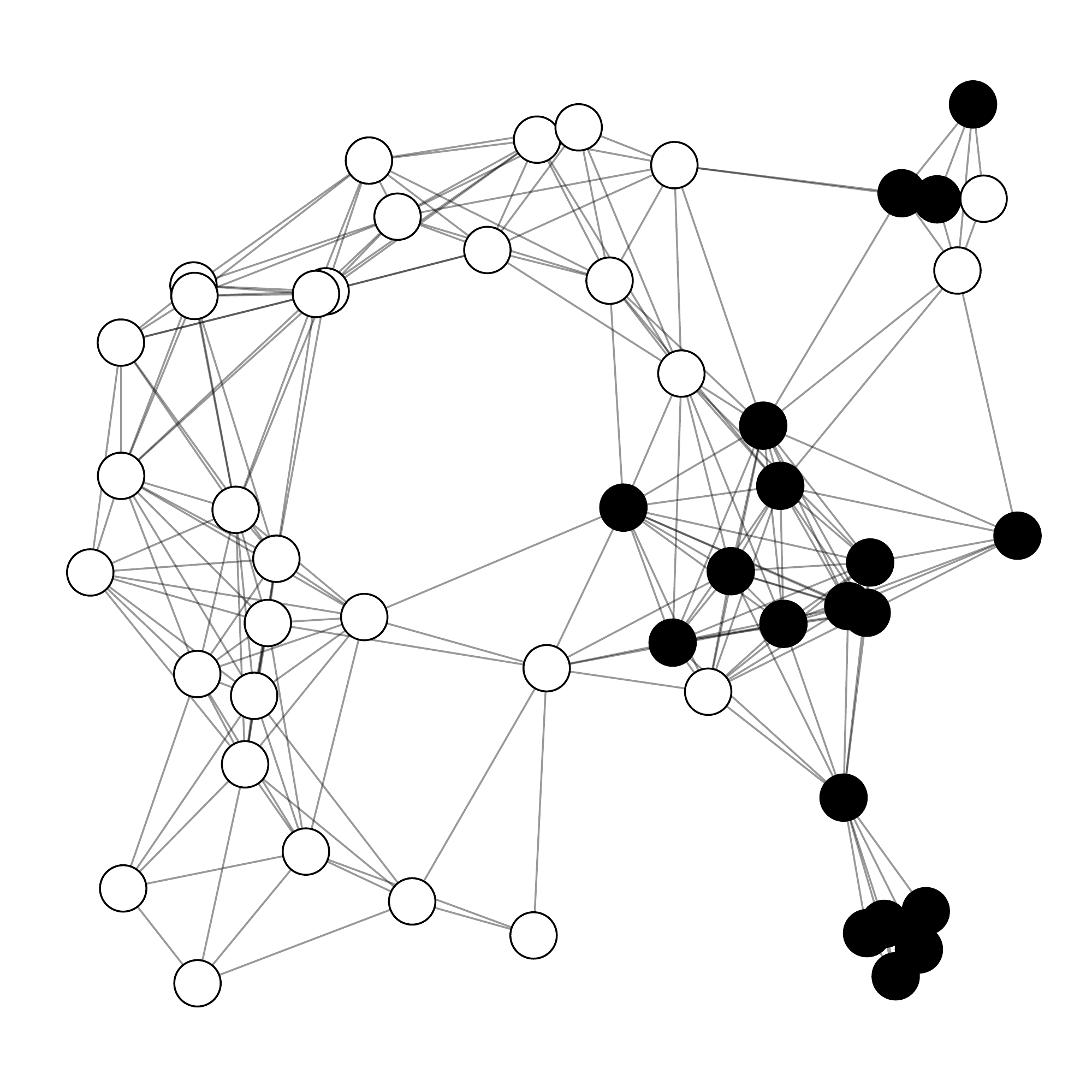} &
      \includegraphics[width=0.18\textwidth]{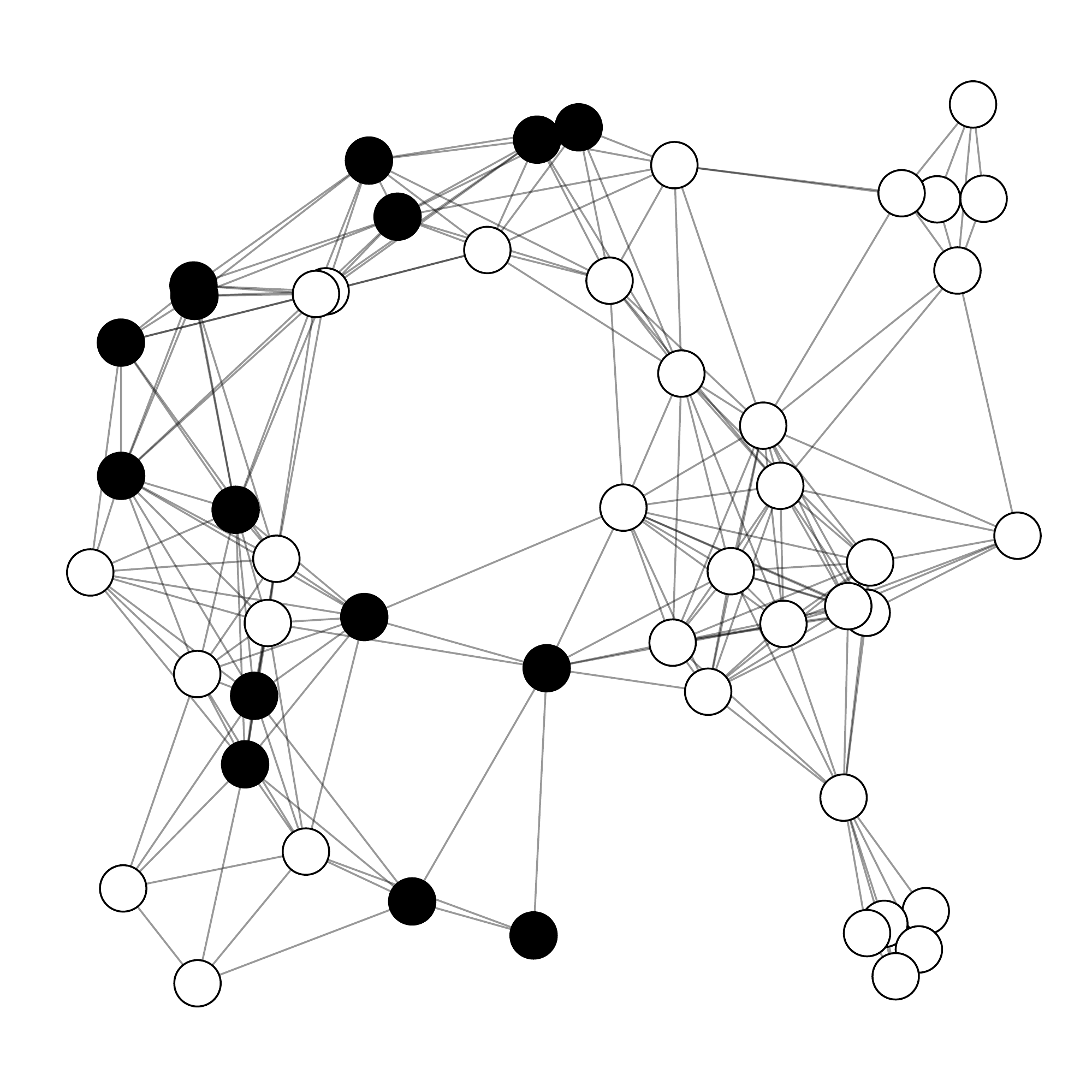} &
      {\includegraphics[width=0.18\textwidth]{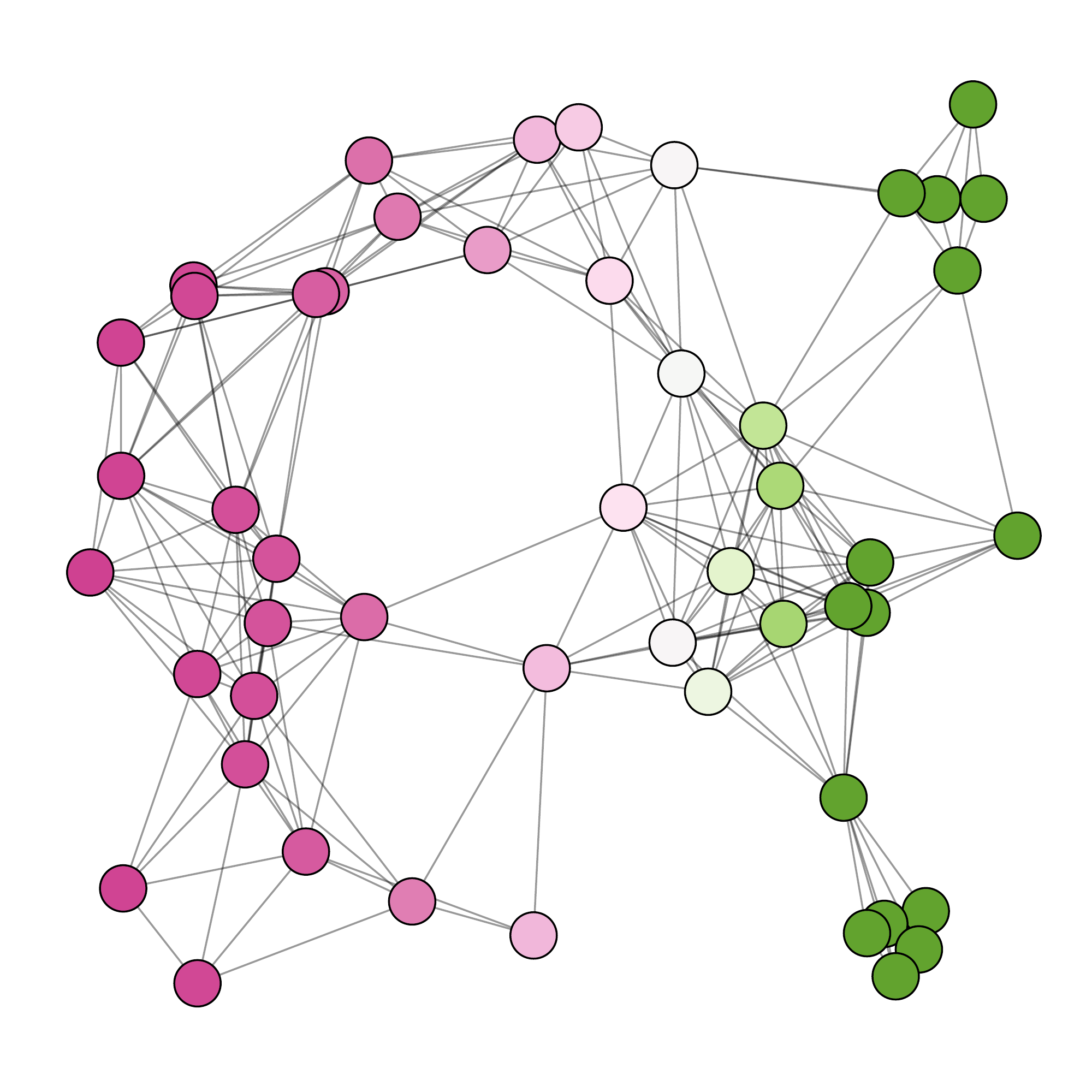}}
      \\
      \multicolumn{2}{c}{$\circ$ = user, $\bullet$ = user sharing an item} \\
      \bottomrule
      \end{tabular}
      }
    \end{center}

    \vspace{-2mm}
        \caption{Abstraction of our proposal.}
    \label{fig:example}
     \vspace{-4mm}
\end{figure}

\begin{figure*}[t!]
\begin{tabular}{ccccc}
  \hspace{-4mm}\includegraphics[width=0.21\textwidth]{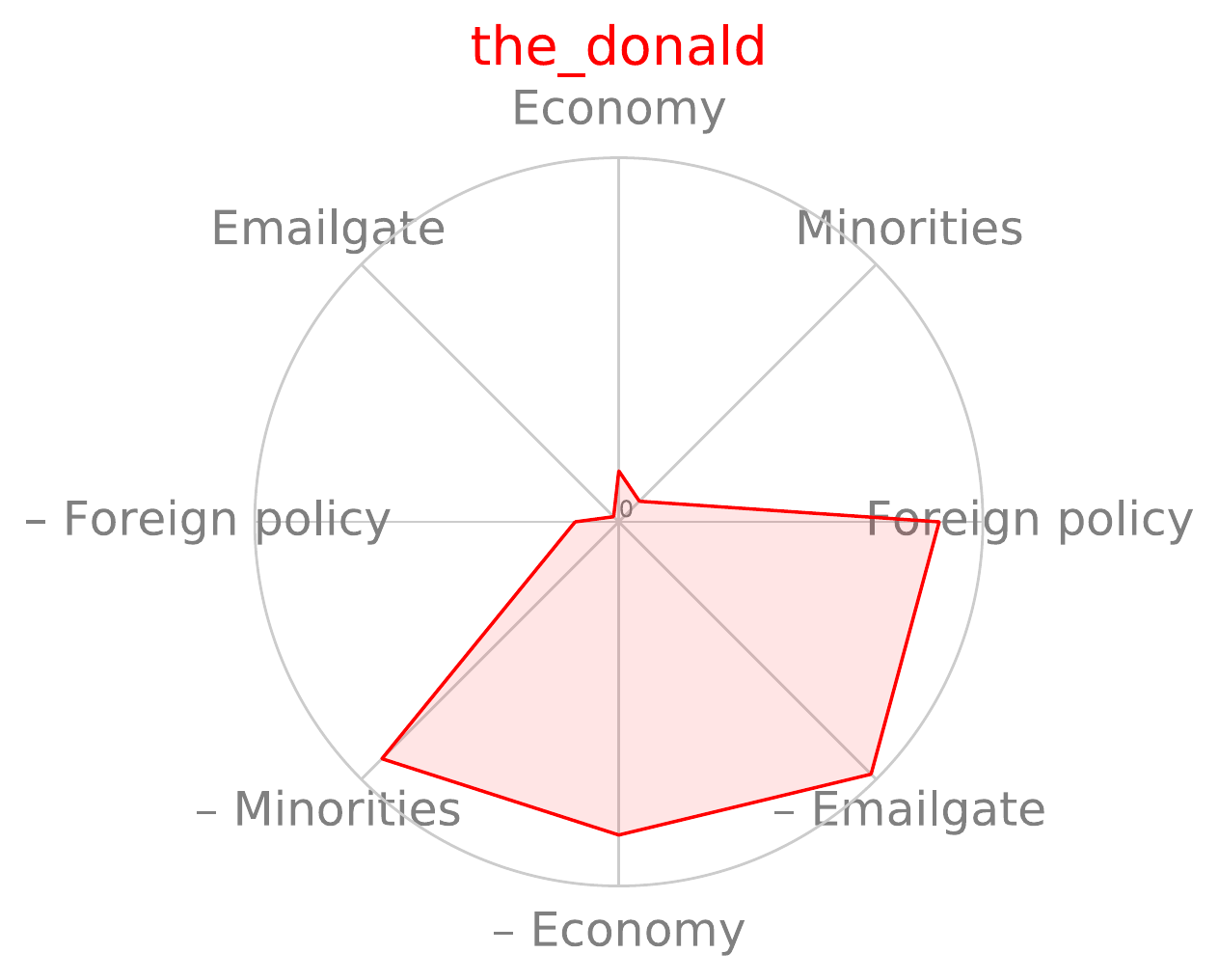} &
  \hspace{-4mm}\includegraphics[width=0.21\textwidth]{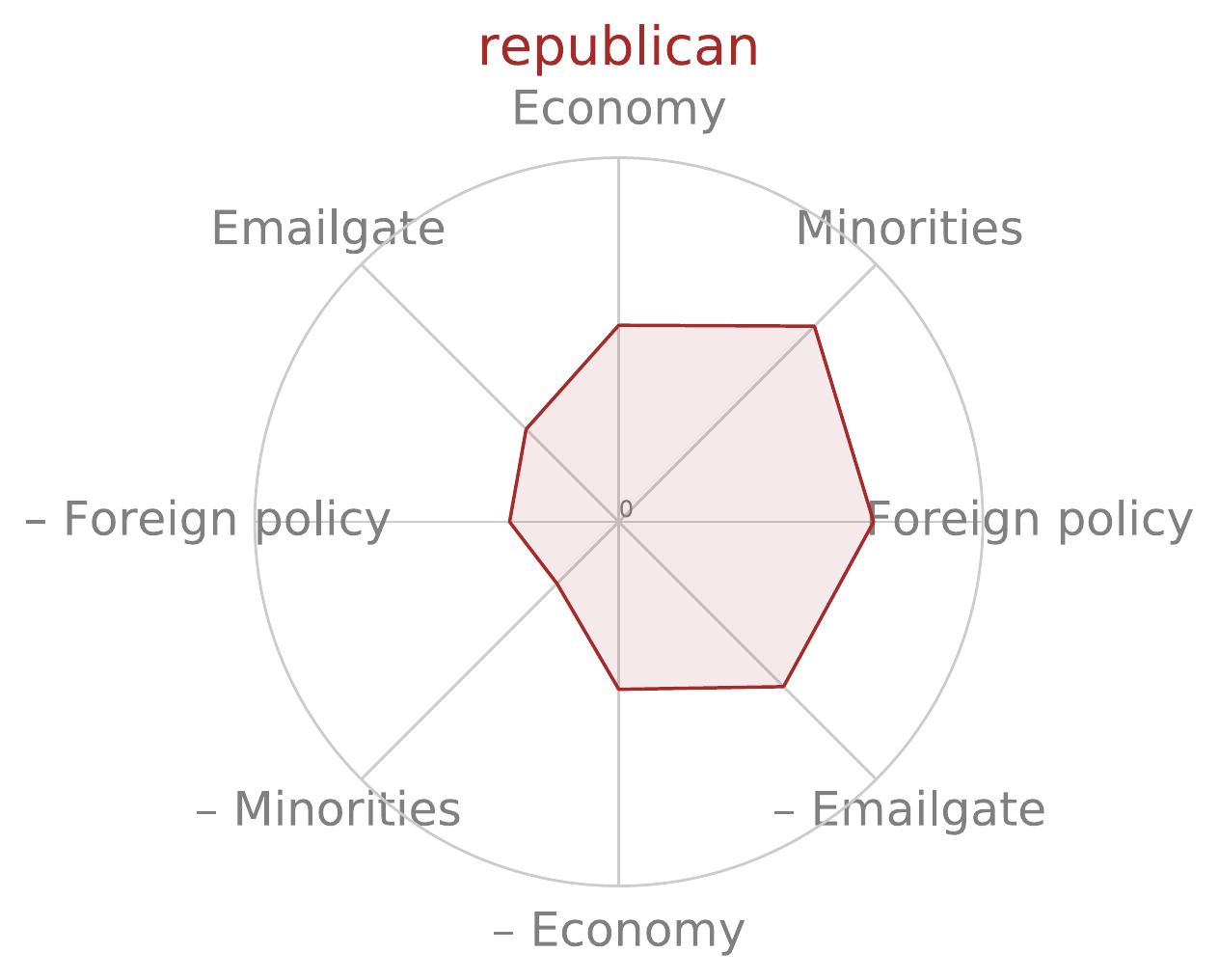} &
  \hspace{-4mm}\includegraphics[width=0.21\textwidth]{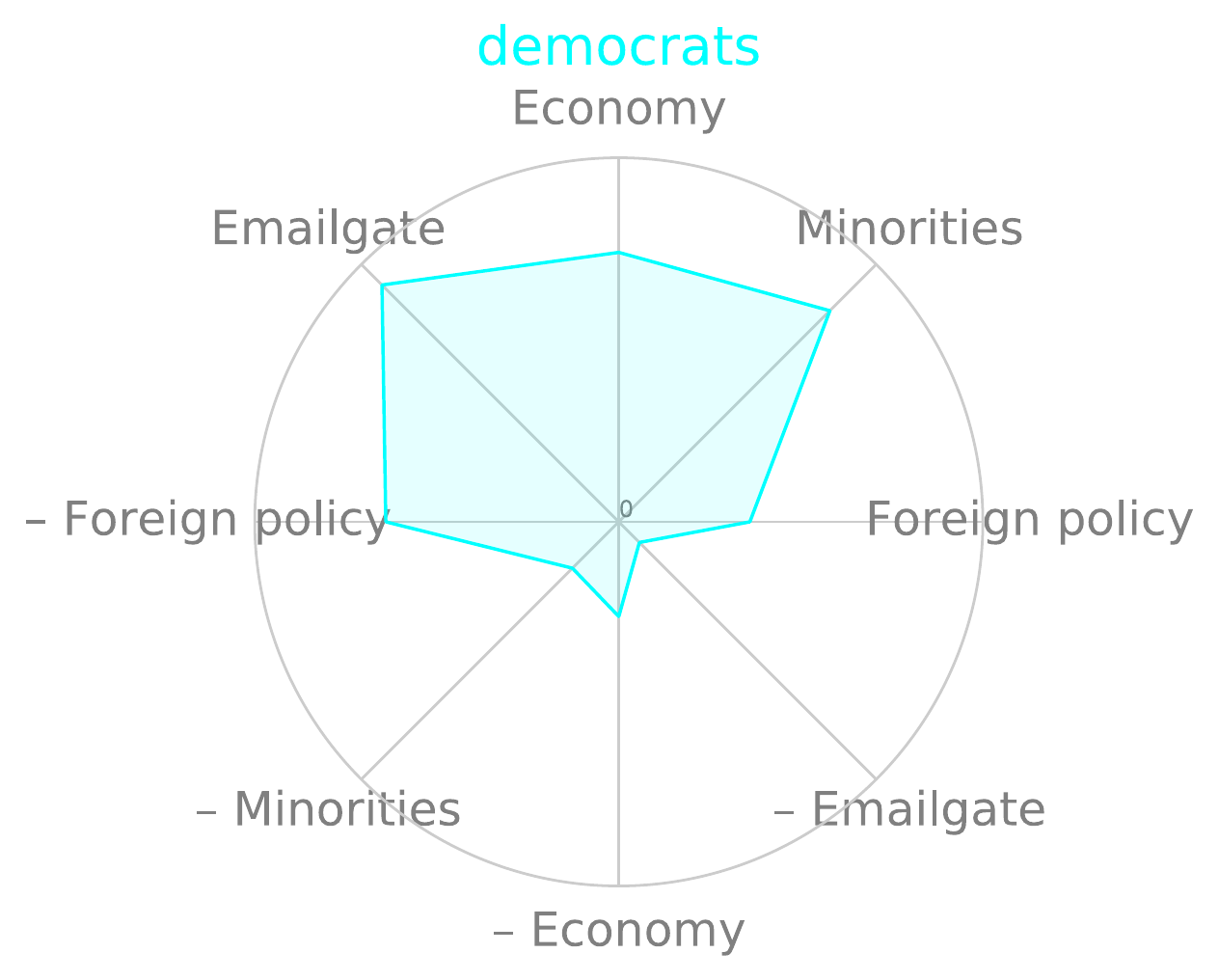} &
  \hspace{-4mm}\includegraphics[width=0.21\textwidth]{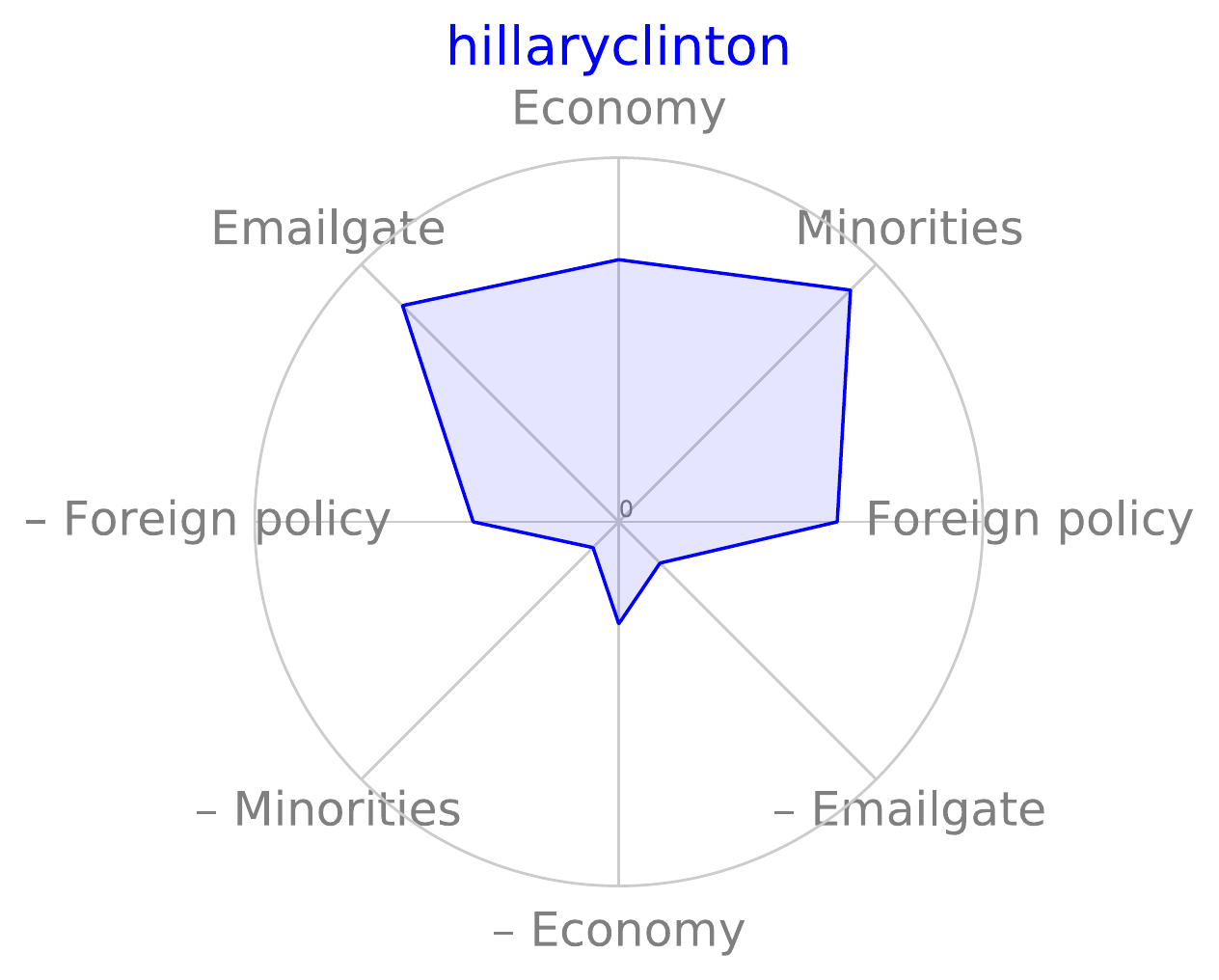} &
  \hspace{-4mm}\includegraphics[width=0.21\textwidth]{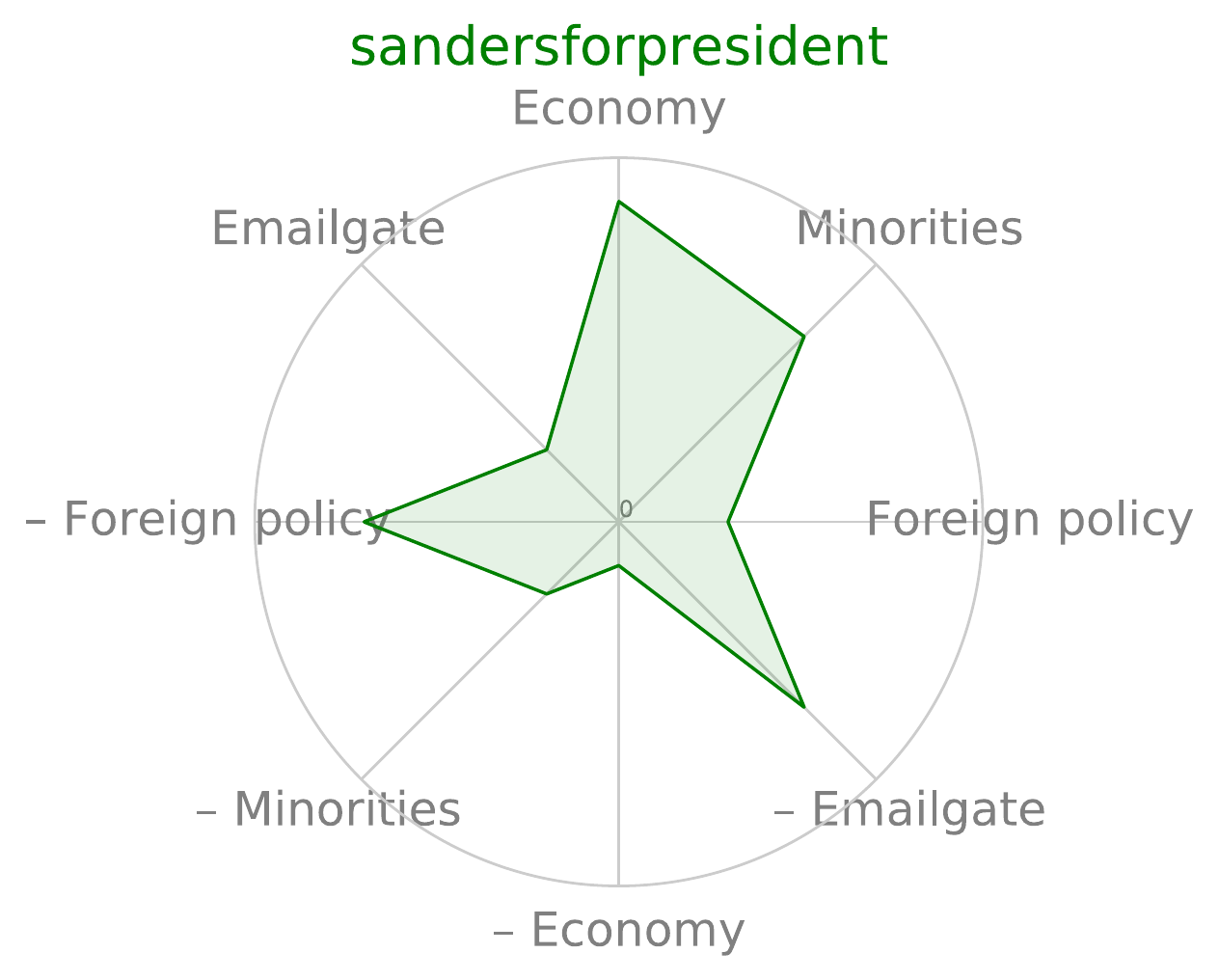}
\end{tabular}
\vspace{-2mm}
\caption{Polarities of some popular subreddits related to 2016 elections, estimated by our method on the Reddit dataset (further details in Sec.~\ref{sec:real-world-subsubsec}). Each of the four dimensions represents one axis of our ideological space.}
\label{fig:radar-plot}
\end{figure*}

In this paper, combining  ideas from different areas -- i.e., network embeddings, information propagation, and opinion mining -- we propose to learn \emph{ideological embeddings} of social media users in a \emph{ multidimensional ideological space} from information cascades. To this end, we propose a stochastic propagation model, dubbed \emph{Multidimensional Ideology-aware Propagation} (MIP) model, formalizing how politically salient content spreads in a social network as a function of the ideological embeddings of users under homophily-driven interactions.
Our model assumes that information propagates from one node in a social network to another if both are interested in the topic and ideologically aligned with each other. Such a scenario can describe for instance information propagations in Twitter, where clusters of retweets have been found to correlate with the political leanings of users~\cite{garimella2018quantifying, barbera2015birds, conover2011political, vaccines}. Accordingly, the ideological embedding of a node is characterized by a topic distribution and polarity for each topic, in a multidimensional space defined by the input topics.
As a result, our embeddings have the nice feature of being \emph{interpretable}, and can be therefore used to make explainable predictions on unobserved behavior.

Figure \ref{fig:example} provides an abstraction of our proposal: the input to our model is the propagation of 4 items, each with a known topic distribution. These observations can be explained by the latent polarities of the users over each topic: in the right part of the figure we indicate the spectrum of polarities on the axis defined by the first topic with colors going from red to blue, and for the second axis from green to purple. These polarities are our output.

As a concrete example, consider the topics: \emph{``economy"}, \emph{``emailgate"}\footnote{``Emailgate'' is the term coined by the media to refer to the controversy around Hillary Clinton's use of a private email server for official public communications. This was used as an argument against her by her opponents in the 2016 U.S. elections.}, \emph{``foreign policy"}, and \emph{``minority rights"}. Assume that $u$ and $v$ are neighbors in a given network (e.g., they are ``friends" or  "follow" each other) but  they disagree on each of these topics. In this case, regardless of how much interested they are in these topics, any content coming from $u$ will hardly be further propagated by $v$ and vice versa.
Alternatively, if $u$ and $v$ are aligned on \emph{``economy"}, then propagation of content about this topic from, e.g., $u$ to $v$ will be likely, as long as $v$ is also exhibiting an interest in the topic.

Notably, the joint explicit modeling of interest and ideological leaning eases the interpretability of nodes' behavior and ultimately the explainability of propagation process. Figure~\ref{fig:radar-plot} provides a preview of some of the results of our method on real-world Reddit data. In particular it reports the polarities learned for some popular subreddits related to 2016 election, along the four axes \emph{``economy"}, \emph{``minorities"}, \emph{``foreign policy"}, and \emph{``emailgate"}. The polarities modeled by our method correspond to expected ones for these subreddits: for instance, along the axis \emph{``emailgate"}, the subreddit \texttt{sandersforpresident} is much closer to subreddits \texttt{republican} and \texttt{the\_donald} than to subreddit \texttt{hillaryclinton}; while on \emph{``economy"} subreddit \texttt{hillaryclinton} is closer to subreddit \texttt{republican} than to \texttt{sandersforpresident}.

\spara{Paper contributions and roadmap.}
The contributions of this paper can be summarized as follows:
\squishlist
\item We propose \model, a topic-aware propagation model of items in a network, where nodes are represented as points in a multidimensional, topic-dependent,  ideological space~(\S\ref{sec:model}).
\item We devise a scalable gradient-based optimization procedure to learn the ideological embeddings that maximize an approximation of the likelihood of a set of information cascades (\S\ref{sec:algorithms}).
\item We provide an extensive empirical evaluation of our proposal on synthetic and real-world datasets and show that our inference algorithm is effective, provides meaningful and interpretable representations for nodes, and can be used to predict unobserved behavior~(\S\ref{sec:experiments}).
\squishend

Next section discusses the most relevant related literature.

\section{Related work}
\label{sec:related}
\textbf{Learning ideological leaning.} Early studies tackling this problem treated it as a classification task and devised methods for predicting the \emph{binary} political alignment of Twitter users~\cite{conover2011predicting,pennacchiotti2011democrats}. Apart from the inherent limitation of one-dimensional political spectrum assumption common to all the existing approaches, \citet{cohen2013classifying} showed that another limitation arose from their need for large amounts of manually annotated data, and their dependence on politically active users.
Some studies looked into content-aware analysis of users' social interactions~\cite{lu2015biaswatch, lahoti2018joint} as we do in our work: none of these approaches, however, exploits information propagation.
In particular, with respect to \citet{lahoti2018joint} our model is (i)~\emph{multidimensional}, since we jointly estimate ideological scores for multiple topics; (ii)~based on \emph{information propagation} rather than content; (iii)~\emph{model-driven}: our estimation algorithm stems from a principled model of how content propagate in a network.

A related research direction, topical stance detection~\cite{DeySK18,ijcai2017-557,Lin+19,sun-etal-2018-stance}, aims at assigning a polarity label to a post towards a specific topic.
We note that these models are only concerned with extracting topics and sentiments from a text corpora; instead, we characterize information propagation by looking at ideological embeddings.

\spara{Influence-driven propagation models.} 
\citet{barbieri2013topic} introduced topic-aware social influence propagation models that take into account the topical interest profiles of the users and the topic distribution of the content that propagates in a social network. Topic-aware social influence modeling is crucial to accurately capture the propagation pattern of content in many applications, ranging from viral marketing and social advertising \cite{AslayBBB14,AslayLB0L15,AslayBLL17} to \emph{who-to-follow} recommendations \cite{BarbieriBM14}. However, for politically salient content, these models are not sufficient to correctly capture the propagation probabilities, as ideological alignment also plays a role.

\spara{Homophily-based models.}
Homophily is a well-known phenomenon is social networks: users tend to interact with users from the same social group~\cite{mcpherson2001birds, himelboim2013birds}.
As such, it is a common assumption in machine learning on social networks~\cite{BarbieriBM13,garimella2018quantifying} and in propagation models~\cite{del2017modeling}.
In this work, we assume that information is more likely to propagate between nodes that, on a specific topic, share the same ideological views.
While the validity of this assumption depends on the specific context and action (replies, follows, etc), it is in general well supported by the current literature.
In the case of Twitter, \emph{retweets} -- the act of resharing a piece of content -- have been shown to follow this pattern very closely:
when a user retweets, it is likely that they share the same ideological background with the source regarding that specific topic~\cite{garimella2018quantifying}.
For example, \citet{conover2011political} found that retweets about U.S. politics had odds ratios higher than one for intra-partisan interactions and lower than $0.1$ for cross-partisan interactions.
\citet{lai2019stance} analyzed users involved in the debate about the 2016 referendum on the reform of the Italian Constitution on Twitter, and  found that 98.6\% of retweets happen between users with the same stance on this topic.
However, this assumption is very general, and it has shown its validity in many other contexts, from blogs~\cite{adamic2005political} to Facebook~\cite{schmidtPolarizationVaccinationDebate2018}.

\spara{Interpretable embeddings.}
We leverage this idea to produce interpretable \emph{ideological embeddings}.
Literature on graph embeddings is vast; we refer to~\citet{goyal2018graph} for a review.
However, most of the techniques from this area offer embeddings that are sub-symbolic, lacking a precise, natural language explanation~\cite{vskrlj2020tax2vec}.
For instance, \citet{fengInf2vecLatentRepresentation2018}~devised a technique to learn embeddings from information cascades: however, their embeddings are not directly interpretable; they do not attempt to model polarities, topics, or opinions.
Some works have tried to deal with the problem of interpretability in graph embeddings. For instance, \citet{idahl2019finding} use an external knowledge base to find interpretable subspaces in a given node embedding space.
\citet{mehta2019stochastic} implement the stochastic block model as a graph neural networks to learn an interpretable embedding for community detection.

To the best of our knowledge, our work is the first to devise an inference algorithm able to extract interpretable embeddings from information cascades;
it is also the first to provide a data-driven model for the interplay between topic-aware opinions and information propagation on social media.

\section{Model}
\label{sec:model}

In this section, we present the \modelfull\ (\model\/ for short).
The model embodies the following set of assumptions:
\begin{enumerate}
  \item Nodes share items, and each item spans a variety of topics.
  \item Each topic corresponds to an \emph{ideological axis}, where each node could be placed.
  \item Nodes adopt the same behavior of the nodes they follow on each item, as long as they agree with each other on the topics of that item.
\end{enumerate}

These assumption are general, they combine aspects from previous works, and are easily applicable to real-world data, as seen in Sec.~\ref{sec:related}.
We next formalize these assumptions in a probabilistic framework.

\spara{Information cascades.}
We consider a directed social graph \mbox{$G = (V,E)$} where $V$ and $E$
denote the set of nodes and edges, respectively, and a directed edge
$(u,v)\in E$ represents $v$ being a \emph{follower} of
$u$. As such, $v$ receives in their timeline the pieces of content (we
call them \emph{items}) shared by $u$ and can be influenced by $u$ to
share further, thus allowing the propagation of
information.
We use the term ``sharing'' here in a loose
sense, i.e., to represent any action which is likely to express an
alignment with the original node, following the assumptions outlined at the beginning of this section. For example, a \emph{retweet} on
Twitter or a \emph{like} on Facebook.
We denote the set of all items as $\items$.
We represent the action of node $u \in \nodes$ sharing an item $i \in
\items$ at time $t \in \mathbb{N}$  with a triplet $(t, i, u)$,
and call such a triplet an \emph{activation}. We assume that no node
adopts the same item more than once, that is, if a node $u$ shares an
item $i$, it becomes \emph{active} on that item, and it will stay so
forever.
In this way, we can focus on observing whether $u$ supports item $i$ or not, and ignore the subtleties regarding multiple sharing behavior; moreover, in some contexts (e.g. a \emph{like} on Facebook) an activation can happen at most once by definition.

We use $\dataset = \{(t, i, u), \dots \}$ to denote the set
of activations observed and $\dataset_i = \{u | (t, i, u)
\in \dataset\}$ to refer to the set of
nodes that became active on $i$ (i.e, the \emph{cascade} of item
$i$).

\spara{Likelihood of a set of cascades.}
Our approach relies on modeling the set $\dataset$ of observed
activations as a result of the stochastic process governed by a set
$\Theta$ of latent factors that span over a multidimensional space,
representing the ideological tendencies of each node. By assuming that
each cascade is independent from the others, the likelihood of
$\dataset$ can be expressed as
$
\mathcal{L}(\dataset; \Theta) = \prod_{i\in \mathcal{I}}
\Pr(\dataset_i|\Theta)
$,
where in turn we can devise $\Pr(\dataset_i|\Theta)$ as
\begin{equation}\label{eq:cascadellk}
\begin{split}
  \Pr(\dataset_i|\Theta) & =  \prod_{u\in \dataset_i} \Pr(\act{u}|\Theta,F_{i,u}) \\
& \cdot \prod_{u \not\in\dataset_i} \left(1-\Pr(\act{u}|\Theta,F_{i,u})\right).
\end{split}
\end{equation}
Here, 
$F_{i,u} = \{v | v \in \dataset_i, (v,u) \in E,
  v  \mbox{ activates before } u\}$
is the ordered set of all in-neighbors of $u$ in cascade $i$.
In other words, the likelihood of a cascade is given by the probability of obtaining the successful and failed activations.
The dependency on
$F_{i,u}$ for modeling the activation probability is fundamental here:
resharing an item depends on those who have already shared
that item. In other words, node $u$ is exposed to $i$ by the nodes followed by $u$ that shared $i$ before.
We consider as activations only those such that $|F_{i,u}| > 0$;
that is, we ignore the initiator of a cascade, as we are only interested in modeling propagation.

\spara{Multidimensional ideological space.}
The core of our approach is then the proper definition of
$\Pr(\act{u}|\Theta,F_{i,u})$ so that it reflects the intuition that activations
only happen within the context of \emph{ideological alignment}.
Since items have different topics which span a low-dimensional space, we assume that

\begin{itemize}
\item an individual only adopts an item if it matches a
  \emph{topic of interest} for them,
\item besides interest, an action is only likely to happen if there is
  an \emph{ideologically alignment on that topic} between the
  individual and the other individuals who shared the item before.
\end{itemize}

These concepts are reflected in the specification of the
model parameters $\Theta$ and consequently in the definition of the activation
probability as follows. First, we assume that each item $i$ exhibits a
multinomial distribution $\boldsymbol\gamma_i$ over $K$  topics (dimensions),
where $\gamma_{i,k}$ represents the relevance of item $i$ to topic
$k$. A node $u$ can exhibit an interest on topic $k$ with
probability $\theta_{u,k}$. The parameter set $\boldsymbol\theta_u$
represents how much $u$ is interested in each of the $K$
topics.
Note that, in principle, a node might even be interested in all the
considered topics, as well as in none of them.

Finally, each of the $K$ topics defines an
\emph{ideological axis}: given a topic $k$, we assume that $u$ can
exhibit either positive or negative leaning within the context of
that topic. The terms ``positive''  and ``negative'' only indicate
that the leanings are opposing each other.
We model the probability of node $u$ exhibiting positive leaning on topic $k$ as
$\phi_{u,k}$ and the probability of negative leaning as
$(1 - \phi_{u, k})$. %

The parameter set $\boldsymbol\phi_u$ represents the set
of all \emph{polarities}, i.e., alignment probabilities for each topic
relative to $u$, and therefore it defines the position of $u$ in the
multidimensional ideological space $[0, 1]^{K}$ defined by the $K$
topics. We  can then express the probability $p(u,v,k)$ that node $u$ is
ideologically aligned  with $v$ on topic $k$ as
\begin{equation}
  p(u,v,k) = \phi_{u,k}\phi_{v,k} + (1-\phi_{u,k})(1-\phi_{v,k}),
\end{equation}
where $\phi_{u,k}\phi_{v,k}$ is the probability that both nodes
exhibit positive leaning on dimension $k$ and, by contrast,
$(1-\phi_{u,k})(1-\phi_{v,k})$ is the probability that they
exhibit negative leaning.
Using this topic-specific alignment probability $p(u, v, k)$, we define
\begin{equation}\label{eq:definition-uv}
  \Pr(\act{u}|\Theta, \act{v}) =
  \sum_k \gamma_{i,k}\cdot \theta_{u,k} \cdot p(u,v,k)
\end{equation}
to be the probability that $u$ will activate on item $i$ given that their predecessor $v$ already did.
In other words, there must be a topic $k$ picked from the topic distribution of the item $\gamma_i$ such that (i) $k$ is of interest to $u$, and (ii) $v$ is ideologically aligned on $k$ with $u$.

Now, we express the probability
$\Pr(\act{u}|\Theta,F_{i,u})$ that a node becomes active on $i$ as a mixture
over all possible activators
\begin{equation}\label{eq:main}
\Pr(\act{u}|\Theta,F_{i,u}) =
\sum_{v \in F_{i,u}} \pi_{i,v}
\Pr(\act{u}|\Theta, \act{v}).
\end{equation}
The term $\pi_{i,v}$ here represent a
prior probability that $v$ causes an activation in $i$.
As such, it must respect $\sum_{v \in F_{i,u}} \pi_{i,v} = 1$; for instance, it can be the uninformative prior $\pi_{i,v} = \frac{1}{ | F_{i,u} | }$. In general, this term adds
additional flexibility to our model, by allowing for a pre-defined
relevance for $v$.
For example,  it can be used to model bias towards
popular nodes, or the initiators of the cascade.

\spara{\model generative model.}
\label{par:generative-model}
We can devise a simple generative stochastic process for data generation that adheres to the aforementioned assumptions.
First, we introduce the hyper-parameters that govern such a generative process.
We define $p$ to control the \emph{polarization}
of the nodes' opinions: high values of $p$ lead to polarity which are
closer to the extremes (i.e., $\phi_{u,k}$ being closer to $0$ or
$1$); lower values lead to polarities closer to neutrality (i.e.,
$\phi_{u,k}$ being closer to $0.5$). We also define $\alpha$ and
$\beta$ to control the generated distribution of the interests. Finally, we define $\mathbf{q}$ as the prior distribution of
the topics of items. Then, given an underlying graph $G=(V,E)$, we take the following steps to generate a propagation dataset.

\medskip

\begin{mdframed}[backgroundcolor=gray!10,roundcorner=10pt]
\begingroup
\fontsize{9.0pt}{10.5pt}\selectfont
\begin{itemize}
  \item Draw interests $\theta_{u,k} \sim \BetaDistribution(\alpha, \beta)$ for each $u \in \nodes$ and topic $k \in \{ 1, \dots, K \}$.
  \item Draw polarities $\phi_{u,k} \sim \BetaDistribution(p^{-1}, p^{-1})$ for each $u \in \nodes$ and topic $k \in \{ 1, \dots, K \}$.
   \item Generate an item $i$:
  \begin{enumerate}
    \item Draw its topic distribution $\gamma_i \sim \Dirichlet(\mathbf{q})$.
    \item Draw an initial activated seed $v$ uniformly at random from~$\nodes$.
    \item For each arc $(v, u) \in E$ s.t. $v$ activated and $u$ has not yet seen the item:
    \begin{enumerate}
      \item Node $u$ sees the item from $v$.
      \item Draw the item topic $k \in \{ 1, \dots, K \}$ according to $\gamma_i$.
      \item $u$ is interested in the topic with probability $\theta_{u, k}$. If it is not interested, then the propagation fails.
      \item If it is interested, draw the attitudes of $u$ and $v$ on $k$ as Bernoulli variables with probability $\phi_{u,k}$ and $\phi_{v,k}$.
      \item If the attitudes are equal, $u$ activates on the item.
    \end{enumerate}
  \end{enumerate}
\end{itemize}
\endgroup
\end{mdframed}

Within the main loop each active node tries to activate its neighbors, and the process stops when no further nodes become active. For simplicity, we presented the model assuming that each active node has an equal chance of activating its neighbors: this corresponds to assuming a uniform prior $\pi$ within Eq.~\ref{eq:main}.

\section{Inference and Learning}
\label{sec:algorithms}

Our goal is to have an algorithm that given a set of items represented
in a $K$-dimensional topic space, is able to estimate the interests
$\boldsymbol\theta_u$ and the polarities $\boldsymbol\phi_u$ of each
node---that is, our ideological embeddings.
We do so by maximizing the likelihood of our model
w.r.t. $\boldsymbol\theta, \boldsymbol\phi$, given all
$\boldsymbol\gamma_i$ and $\dataset_i$.

We can easily see the problem of maximizing the likelihood we defined
in the previous section as a learning problem, in which we predict
whether a certain node will activate or not.
To obtain a scalable learning algorithm, we would like to treat each possible propagation
of an item from $v$ to $u$ as an independent example.
To do so, we resort to an approximation of the
likelihood in Eq.~\ref{eq:cascadellk}. First, for the active users, we notice
that, by Jensen inequality:
\begin{equation*}
    \log \Pr(\act{u}|\Theta,F_{i,u}) \geq  \sum_{v\in F_{i,u}} \pi_{i,v}\left[\log \Pr(\act{u}|\Theta, \act{v}) \right].
\end{equation*}
By virtue of this inequality, we can consider the prior $\pi_{i,v}$ as a
predefined sampler that selects users from $F_{i,u}$ and approximates the
activation probability according to them. In practice, we can focus on
a subset of predefined potential activators, and measure the
probability of alignment to them.

Different choices of which nodes to sample suit different contexts.
For example, considering only the first
activator is appropriate in some real cases, where the ideological
alignment is likely to happen mostly between the creator of the item
and whoever adopts it, more than between followers. %
In our algorithm, however, we wish to treat  as equal every pair $u, v$ among those that reshared an item $i$.
This assumption comes from the empirical observation that political
communities tend to form groups with a high
homophily, often described as echo chambers~\cite{del2016echo}; these appear in retweet networks as clusters with a homogeneous opinion~\cite{garimella2018quantifying, barbera2015birds, conover2011political, vaccines}.
Therefore, we expect aligned nodes to activate \emph{collectively} -- i.e., if
$u$ is aligned with a neighbor who shared $i$, they are also probably
aligned with another neighbor who shared the same item.

Based on this practical consideration, we further choose to
approximate the negative terms as
\begin{equation*}
  1 - \Pr( \act{u}| \Theta,F_{i,u}) \approx
  \prod_{v \in F_{i,u}} \left( 1-\Pr( \act{u}| \Theta, \act{v})\right).
\end{equation*}
As a matter of fact, since ideological communities tend to be homogenous, then if $u$
is \emph{not} aligned with a neighbor who shared $i$, we expect them
in practice to be unaligned with \emph{all} their neighbors who shared
the same item.
This approximation allows us first to better learn from real-world
propagations, clustering together users who reshared an item and
separate those who did not.
Secondly, it allows us to factorize the likelihood and hence to obtain a scalable algorithm. We can therefore rewrite the overall log-likelihood as
 \begin{equation*}\small
  \begin{split}
    \log \Pr(\dataset_i,|\Theta) \approx & \sum_{u\in \dataset_i} \sum_{v\in F_{i,u}} \log\left(\sum_k \gamma_{i,k}\cdot
                              \theta_{u,k} \cdot p(u,v,k)\right) \\
   &   + \sum_{u \not\in \dataset_i}\sum_{v\in
                              F_{i,u}}\log\left(1 - \sum_k \gamma_{i,k}\cdot
                              \theta_{u,k} \cdot p(u,v,k)\right)%
                              .\\
   \end{split}
 \end{equation*}

Thanks to this approximation, we can employ stochastic gradient
descent in the following way. The basic instance of the learning
problem is an example $x,y$ where
$x \equiv \langle \boldsymbol\gamma_i, v, u  \rangle$ and $y = 1$ iff $u
\in \dataset_i$ and $0$ otherwise, given that $v \in F_{i,u}$.
In other words, from the topic distribution of an item and one of the
nodes active on it, we try to predict whether one of its
followers will activate or not.  Then, given a sequence of examples
$X$ sampled from the set of observed
activations $\dataset$, maximizing the likelihood is
equivalent to minimizing the log loss of each example:
$$ \phi, \theta = \argmin_{\phi, \theta} -\sum_{x, y \in X} y \, \log
\left( Pr(x) \right) + (1 - y) \, \log \left( 1 - Pr(x) \right) $$
where $Pr(x)$ is the likelihood of a single example given the latent
variables $\phi$ and $\theta$.

\begin{algorithm}[t]
    {
    \caption{\model inference algorithm.}
    \label{alg:inference}
    \begingroup
    \fontsize{8.0pt}{9.0pt}\selectfont
    \begin{flushleft}
    \algorithmicrequire \; $G=(V, E)$; items $i \in \items$ with %
      topics $\gamma_i \in \mathbb{R}^K$, activations $\dataset_i \subseteq V$.
    \algorithmicensure \; Polarities $\phi_u$ and interests $\theta_u$ for all $u \in \nodes$.
    \begin{algorithmic}[1]
        \State Initialize $\phi$ and $\theta$ as $|V| \times K$ matrices.
        \For{number of epochs}
          \For{$i \in \items$}
            \For{$v \in \dataset_i$} \label{line:seed-sampling}
              \For{$u \in \{ u \in \dataset_i | (v, u) \in E \}$ }
                \State Update $\phi, \theta$ by ascending the gradient:
                \begin{equation*}
                  \nabla_{\phi, \theta}
                  \log \left( \sum_k \gamma_{i, k} \cdot \theta_{u, k}
                    \cdot p(u, v, k) \right)
                \end{equation*}
              \EndFor
              \For{$u \in \sample \left( \{ u \notin \dataset_i | (v,u) \in E \} \right) $ } \label{line:negative-sampling}
                \State Update $\phi, \theta$ by ascending the gradient:
                \begin{equation*}
                  \nabla_{\phi, \theta}
                  \log \left( 1 - \sum_k \gamma_{i, k} \cdot
                    \theta_{u, k} \cdot p(u, v, k) \right)
                \end{equation*}
              \EndFor
            \EndFor
          \EndFor
        \EndFor
        \State{Return $\phi$ and $\theta$.}
    \end{algorithmic}
  \end{flushleft}
  \endgroup
}
\end{algorithm}

The learning procedure is outlined in Algorithm~\ref{alg:inference}.
In real data sets, non-activated nodes largely outnumber the activated
ones for a given item. For this reason, we use negative undersampling
(line~\ref{line:negative-sampling}) to reduce the number of negatives.
In practice, we undersample negative examples at random so that they are two times the number of positives.
In experiments, this implies an average precision of $0.33$ in expectation when predictions are performed randomly.

We also found that considering a random sample of $\dataset_i$ of fixed size (line~\ref{line:seed-sampling}) seems to work well in practice.
For gradient descent, we employed AdaGrad~\cite{duchi2011adaptive} with a linearly decreasing learning rate (stopping at $0.01$) to avoid undesirable local minima.

\begin{figure*}[t]
    \centering
    \begin{subfigure}[b]{0.22\textwidth}
        \centering
        {\footnotesize \begin{tabular}{rrr}
\toprule
 Num. items &  AUC ROC &  Avg. Prec. \\
\midrule
       1000 &    0.570 &       0.497 \\
      10000 &    0.754 &       0.717 \\
     100000 &    0.826 &       0.797 \\
\bottomrule
\end{tabular}
 }
        \includegraphics[width=\textwidth]{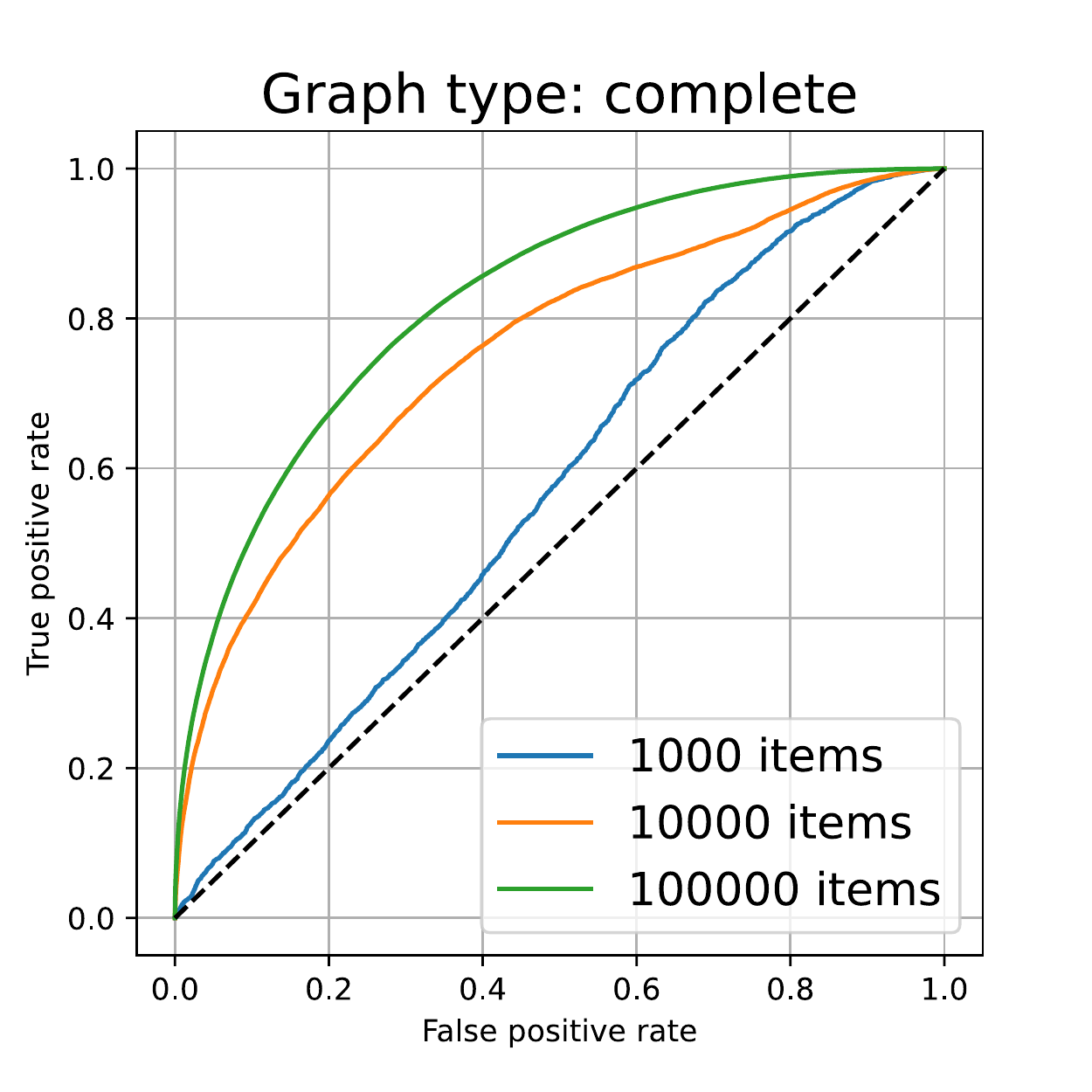}
        \vspace{-7.5mm}
        \caption{}
        \label{fig:synth-num-items-roc}
    \end{subfigure}
    \quad
    \begin{subfigure}[b]{0.22\textwidth}
        \centering
        {\footnotesize \begin{tabular}{rrr}
\toprule
 Num. items &  AUC ROC &  Avg. Prec. \\
\midrule
       1000 &    0.607 &       0.543 \\
      10000 &    0.773 &       0.733 \\
     100000 &    0.840 &       0.815 \\
\bottomrule
\end{tabular}
 }
        \includegraphics[width=\textwidth]{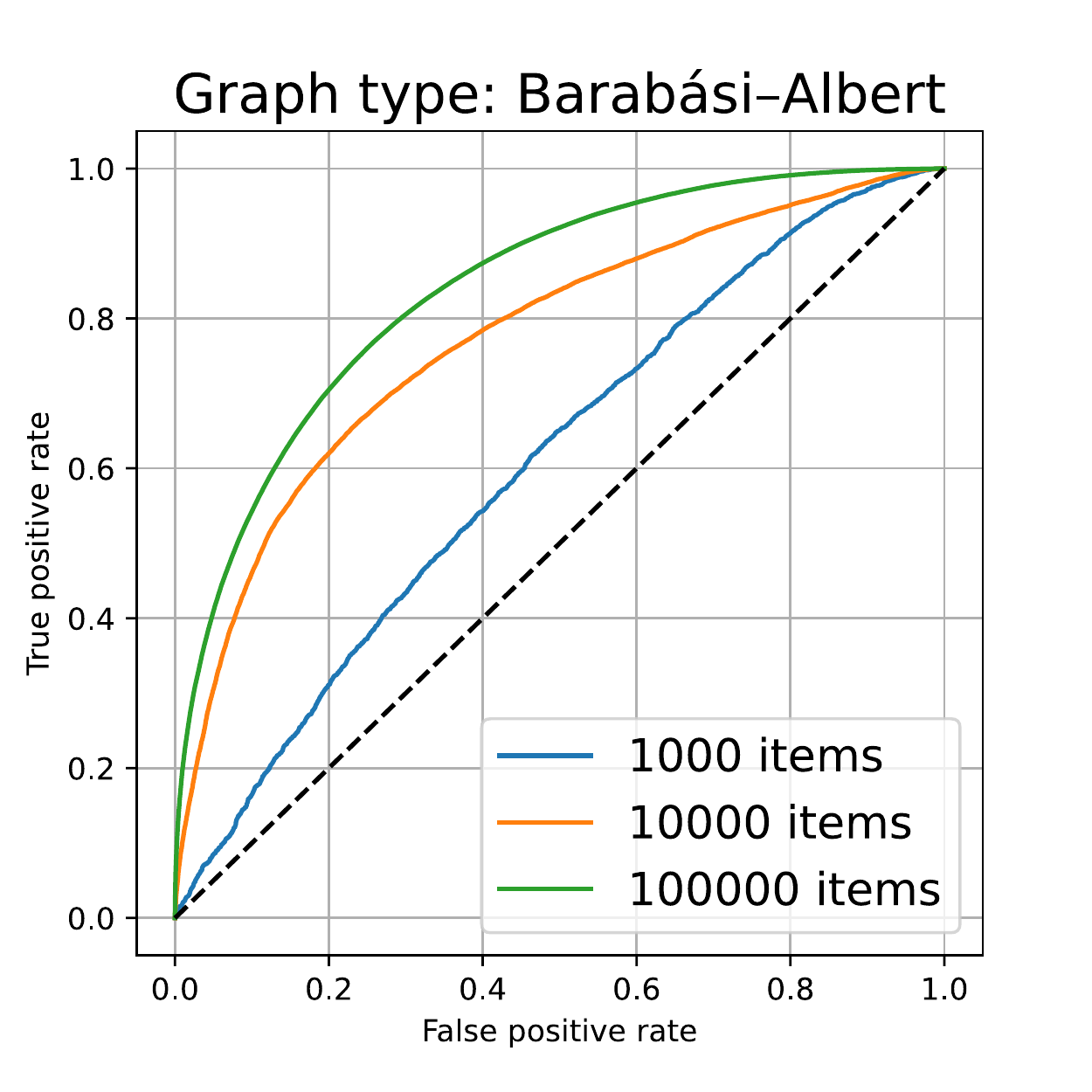}
        \vspace{-7.5mm}
        \caption{}
        \label{fig:synth-num-items-roc-barabasi}
    \end{subfigure}
    \quad
    \begin{subfigure}[b]{0.22\textwidth}
        \centering
        {\footnotesize \begin{tabular}{rrr}
\toprule
 $p$ &  AUC ROC &  Avg. Prec. \\
\midrule
   1 &    0.609 &       0.530 \\
   4 &    0.754 &       0.717 \\
  16 &    0.889 &       0.870 \\
\bottomrule
\end{tabular}
 }
        \includegraphics[width=\textwidth]{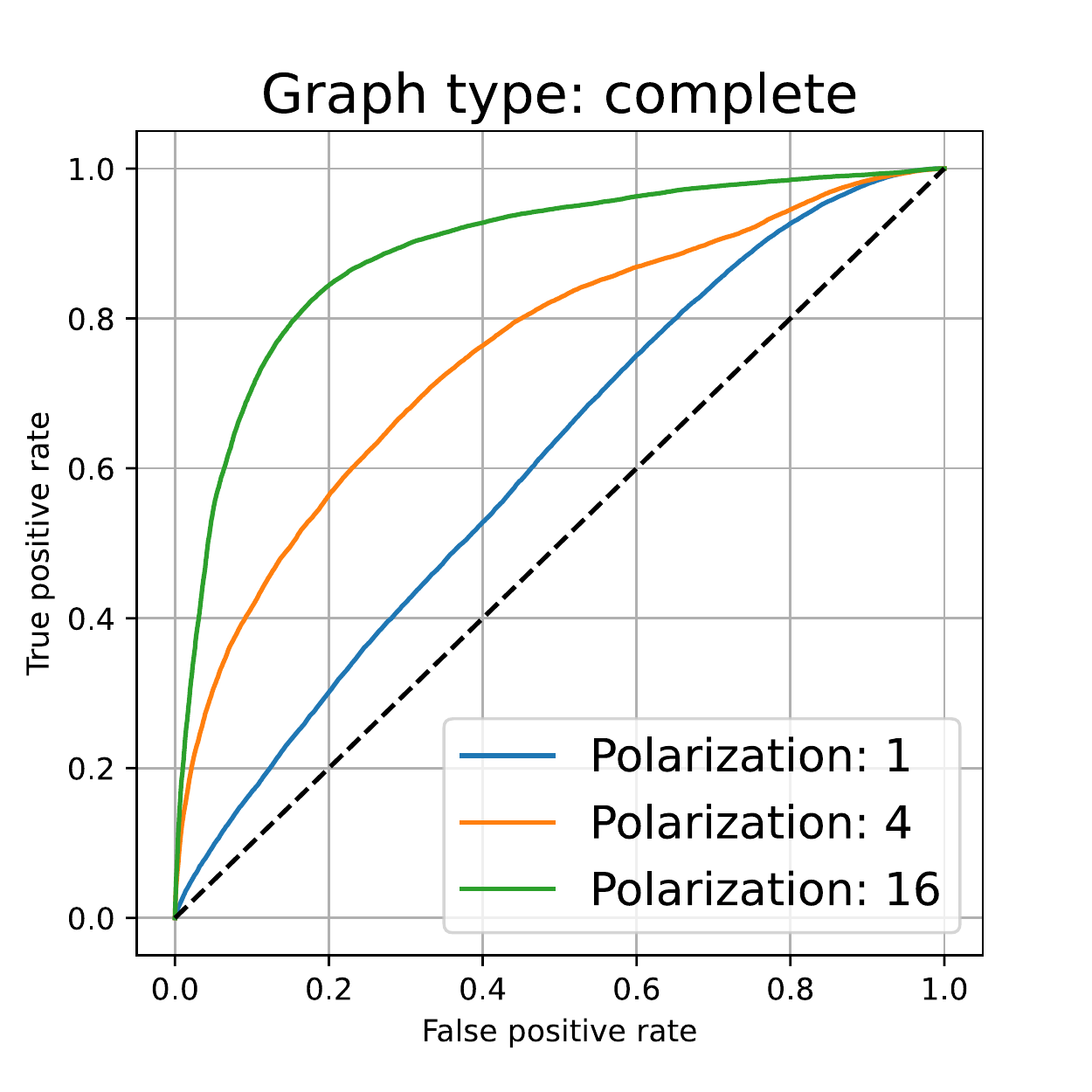}
        \vspace{-7.5mm}
        \caption{}
        \label{fig:synth-polarization-roc}
    \end{subfigure}
    \begin{subfigure}[b]{0.22\textwidth}
        \centering
        {\footnotesize \begin{tabular}{rrr}
\toprule
 $p$ &  AUC ROC &  Avg. Prec. \\
\midrule
   1 &    0.601 &       0.534 \\
   4 &    0.773 &       0.733 \\
  16 &    0.884 &       0.857 \\
\bottomrule
\end{tabular}
 }
        \includegraphics[width=\textwidth]{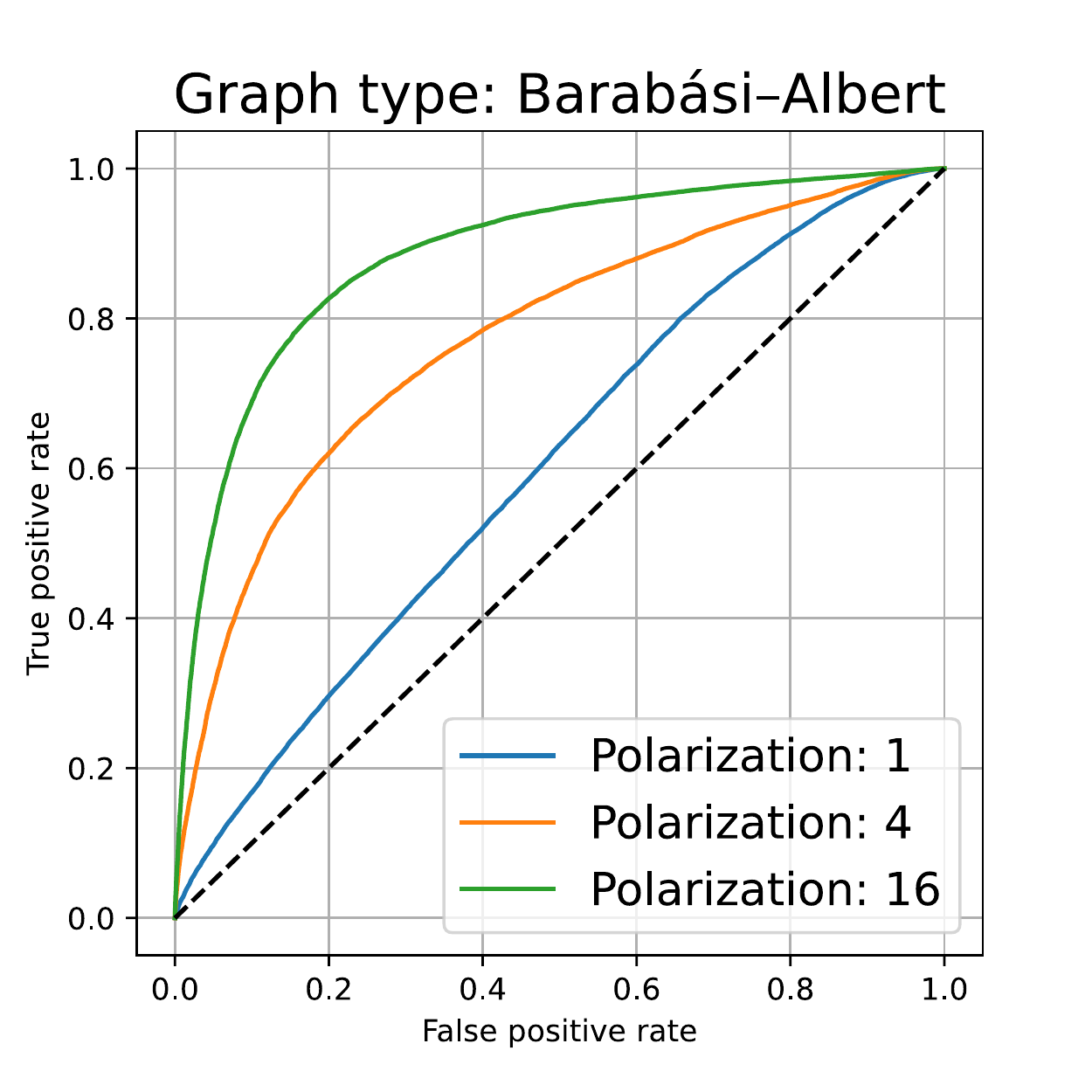}
        \vspace{-7.5mm}
        \caption{}
        \label{fig:synth-polarization-roc-barabasi}
    \end{subfigure}
    \vspace{-4mm}
\caption{Accuracy of the inference algorithm on synthetic data. %
Predictions are increasingly more accurate with the increasing number of items (Figure~\ref{fig:synth-num-items-roc} and \ref{fig:synth-num-items-roc-barabasi}) and with more polarized opinions (Figure~\ref{fig:synth-polarization-roc} and \ref{fig:synth-polarization-roc-barabasi}). }
    \vspace{-3mm}
\end{figure*}

\section{Experimental Assessment}
\label{sec:experiments}

We conduct an extensive empirical evaluation of the proposed model and the inference algorithm on synthetic and real-world datasets. Our goal is to answer to the following research questions:

\begin{itemize}
  \item \textbf{RQ1}. By assuming that propagations meet the assumptions of section~\ref{sec:model}, can our algorithm infer accurate latent ideological embeddings that justify such propagations? Under which conditions? (Section~\ref{sec:in-vitro})
  \item \textbf{RQ2}. Are the predictions from our algorithm explainable? More precisely: are the embeddings inferred by our algorithm from a real-world scenario \emph{interpretable} and \emph{meaningful}, compared to well-known facts about ideological positioning of political groups? (Section~\ref{sec:reddit-in-depth})
  \item \textbf{RQ3}. In real-world scenarios, can ideological embeddings be used to predict unobserved behavior?
  How does their predictive power compare to less interpretable \mbox{state-of-the-art} embeddings?
  (Section~\ref{sec:real-world})
\end{itemize}

In order to foster reproducibility, we publicly release all the data and code necessary to reproduce our experiments.\footnote{\url{https://github.com/corradomonti/ideological-embeddings}}

\subsection{Experiments on synthetic data}
\label{sec:in-vitro}

To answer the first question, we wish to verify that our inference algorithm can make correct predictions whenever activations follow the assumptions outlined in Section~\ref{sec:model}.
To this end, we conduct experiments on multiple synthetic datasets in which activations are generated according to our model. %
We show that Algorithm~\ref{alg:inference}, despite the approximations we assumed to make it scalable, is consistent with the generative process proposed in Section~\ref{par:generative-model}.
Furthermore, we study under which conditions the predictions made by the algorithm are accurate: it turns out that polarization affects greatly its accuracy, as does the number of items; the topology of the underlying graph, instead, seems to have less impact.

\spara{Experimental setting.} In all our experiments, we consider the items as divided into training ($90$\%) and test ($10$\%) sets. We run our inference algorithm on the training set to estimate the values of the  variables $\phi$ and $\theta$ which are then used to compute the likelihood of activations in the test set. In particular, we consider the following prediction task: given that a node $v$ has activated on an item $i$ with topic distribution $\gamma_{i}$, what is the probability that node $u$ will also activate on item $i$?

To generate the synthetic dataset, we follow the procedure described in Section~\ref{par:generative-model} with the following hyper-parameters.
We set $K=4$ topics.
We set $\alpha = 0.9$ and $\beta = 0.1$, in order to have a high ($92\%$) probability
of users having topic-aware interests $\theta_{u,k}$ greater than
$0.5$. As a prior for items, we set $q=1/8$ for all the topics.
We use %
two different graphs $G$, both with $|\nodes| = 100$ nodes: (i)~a~complete graph, with density $1$; (ii)~a~graph generated by the Barab{\'a}si-Albert network model, %
with $M=10$ and density $0.18$.

\spara{Results.} We first compare the accuracy of the inference process when varying the number of items propagated through the network. To this end, we set $p=4$ and perform inference with $|\items| \in \{1000, 10\,000$, $100\,000\}$ items.
We report the results in Fig.~\ref{fig:synth-num-items-roc}~and~\ref{fig:synth-num-items-roc-barabasi}.
As expected, the inference accuracy increases with the number of items since every new item introduces a new cascade to the sample, improving the accuracy of parameter estimations.
Surprisingly, the results are similar for both the complete graph and the Barab{\'a}si-Albert network model, suggesting that the topology does not seem to affect the performance of our algorithm.

We then measure the inference accuracy with regards to the distribution on the polarization of the individuals. To this end, we set $|\items| = 10\,000$ and perform inference on the datasets generated by setting the hyper-parameter $p \in \{1, 4, 16\}$.
We report the results in Fig.~\ref{fig:synth-polarization-roc}~and~\ref{fig:synth-polarization-roc-barabasi}. We can clearly see that our inference algorithm performs better in highly polarized scenarios (i.e., higher values of $p$) when items induce great controversy.
This result shows that more polarized nodes lead to different behavior in the propagation of the content, making it easier for the inference algorithm to distinguish the leaning of the nodes.
As a result, we see that the polarization has a profound effect on the quality of the estimations.
Overall, we can conclude that our inference algorithm is effective in estimating polarities and interests in order to make predictions about new unseen activations.

\subsection{Interpretable Embeddings}
\label{sec:reddit-in-depth}
In this section, we answer RQ2, that is: are the predictions made by our algorithm explainable? Or, in other words, is our algorithm able to extract \emph{interpretable} embeddings when applied to a real-world dataset? Is the interpretation of those embeddings in line with known facts about a real world dataset?
We operationalize these questions by employing a dataset based on well-studied~\cite{massachs2020roots, soliman2019characterization, mills2018pop} communities from Reddit.
We will show how the embeddings we find correspond to known political positions of these communities.
We refer to Section~\ref{sec:real-world} for an analysis of the predictive power of such embeddings.

\spara{Reddit dataset.}
The real-world dataset we consider in this part is extracted by crawling the
social news aggregation website Reddit. Here, we aim at capturing how
different Reddit communities (\mbox{\emph{subreddits}}) share political news items.
In fact, we focus on subreddits as nodes, while we ignore individual users: in this way, we can use their known political position as a validation of the interpretability of our embeddings.

To this end, we interpreted an item $i$ to be a specific URL posted on Reddit.
Each subreddit community is a node $u$ that can propagate (i.e. post) an item.
We consider subreddits as nodes of a complete graph (i.e., each subreddit can reshare news from all the others).
We say that a node $u$ is activated on an item $i$ when the URL
that corresponds to item $i$ is posted on the subreddit~$u$.

We identified $50$ subreddit communities that are most similar to \texttt{r/politics} (the main community for U.S. politics) in terms of their cosine similarity over users.%
We then collected the URLs that have been shared by at least $5$ of these $50$ subreddits between the years $2011$ and $2018$, obtaining a set of $22\,047$ items.

We are therefore assuming that this dataset fits the assumptions we outlined at the beginning of Section~\ref{sec:model}: i.e., each posted item spans a variety of topics; subreddits share an item already shared on another subreddit if they are interested in it and if the two communities are aligned on the topics of that specific item.

\begin{figure*}[t]
\begin{minipage}[c]{.4\linewidth}
\includegraphics[height=.95\linewidth]{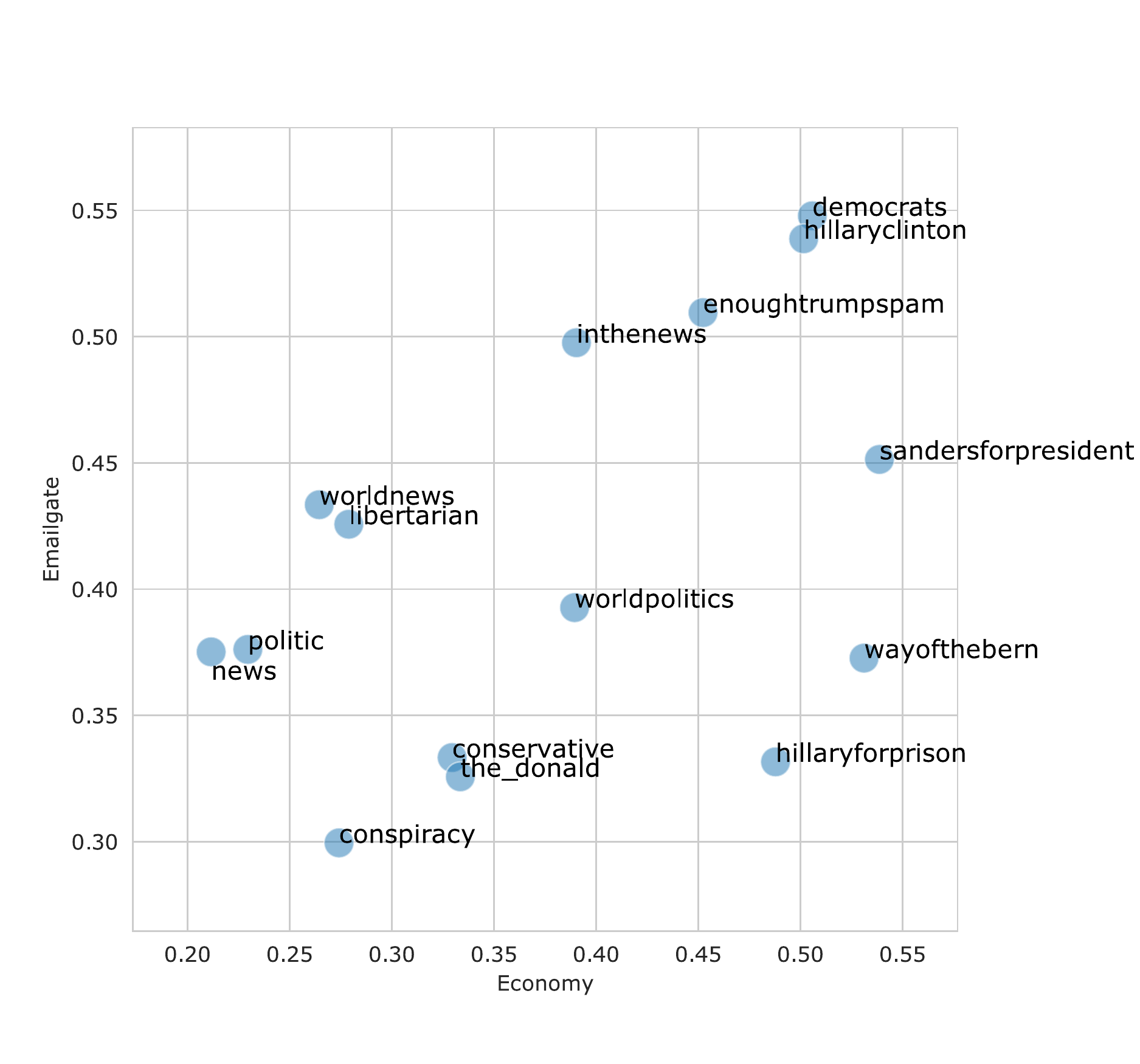}
\end{minipage}
\begin{minipage}[c]{.4\linewidth}
\includegraphics[height=.95\linewidth]{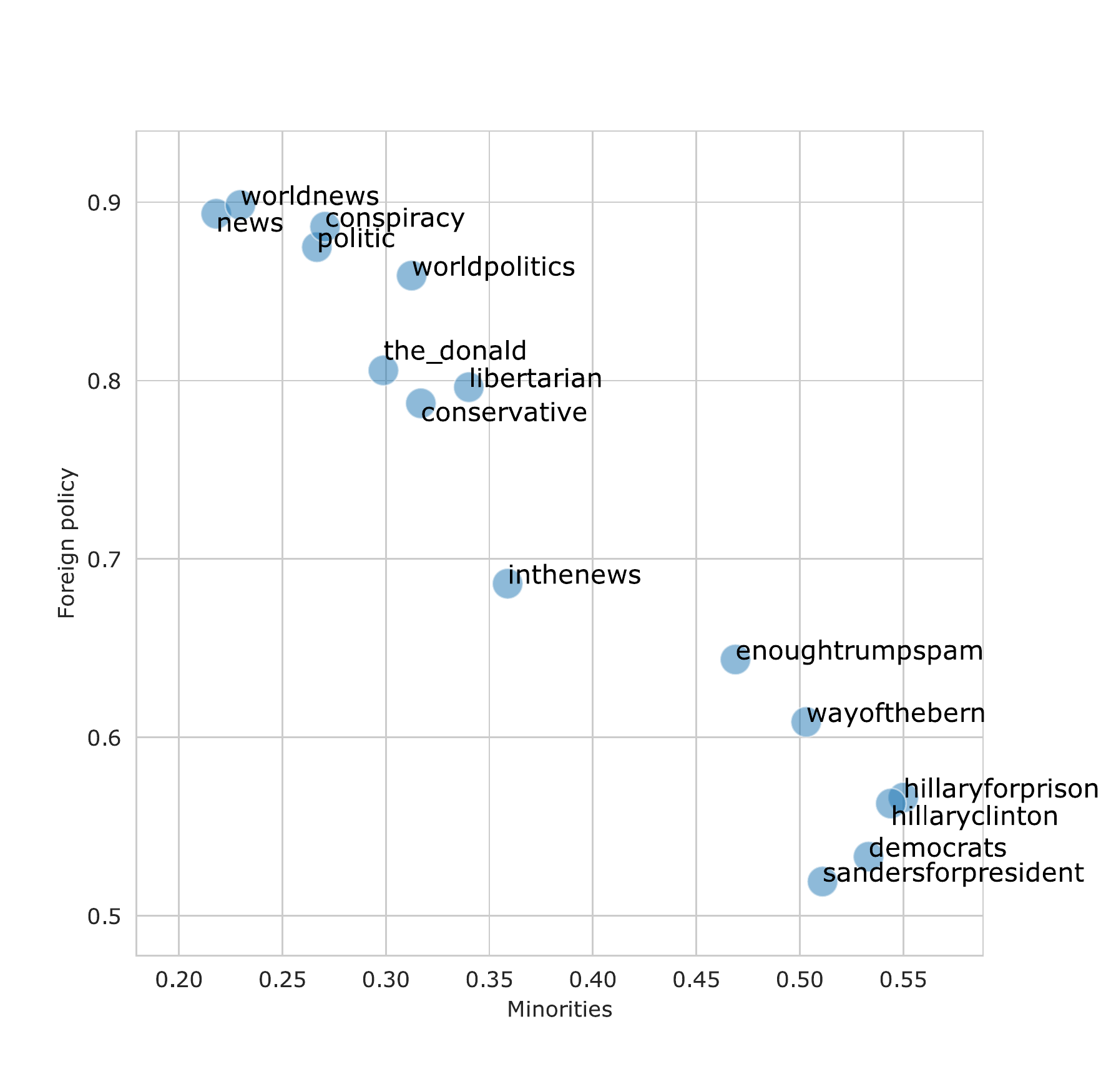}
\end{minipage}
\begin{minipage}[c]{.14\linewidth}
      \fontsize{5.5}{5.5} \selectfont
		  \begin{tabular}{p{8em}p{4em}}
  \toprule
             \emph{Subreddit} &  \textbf{\textsf{Economy}} \\
  \midrule
         libertarian &  0.278960 \\
          the\_donald &  0.333547 \\
      hillaryclinton &  0.501605 \\
           democrats &  0.505838 \\
        wayofthebern &  0.531154 \\
 sandersforpresident &  0.538688 \\
\cline{1-2}
\end{tabular}
 \\ \vspace{.2mm}
      \begin{tabular}{p{8em}p{4em}}
\toprule
           \vspace{.3mm} \emph{Subreddit} &  \textbf{\textsf{Foreign policy}} \\
\midrule
 sandersforpresident &        0.519180 \\
           democrats &        0.533123 \\
      hillaryclinton &        0.562925 \\
        wayofthebern &        0.608667 \\
         libertarian &        0.796325 \\
          the\_donald &        0.805734 \\
\cline{1-2}
\end{tabular}
 \\ \vspace{.2mm}
      \begin{tabular}{p{8em}p{4em}}
  \toprule
             \emph{Subreddit} &  \textbf{\textsf{Minorities}} \\
  \midrule
          the\_donald &    0.298800 \\
         libertarian &    0.340077 \\
        wayofthebern &    0.503109 \\
 sandersforpresident &    0.511015 \\
           democrats &    0.533224 \\
      hillaryclinton &    0.544002 \\
\cline{1-2}
\end{tabular}
 \\ \vspace{.2mm}
      \begin{tabular}{p{8em}p{4em}}
  \toprule
             \emph{Subreddit} &  \textbf{\textsf{Emailgate}} \\
  \midrule
          the\_donald &   0.325649 \\
        wayofthebern &   0.372663 \\
         libertarian &   0.425686 \\
 sandersforpresident &   0.451363 \\
      hillaryclinton &   0.538760 \\
           democrats &   0.547866 \\
\cline{1-2}
\end{tabular}

\end{minipage}
\caption{Estimated polarities of the most popular subreddits (2016 U.S. presidential) on several axes.}
\label{fig:reddit-polarity-plot}
\vspace{-3mm}
\end{figure*}

\spara{Ideological axes.}
To define the ideological axes of our model we explored the dataset using \texttt{doc2vec}~\cite{le2014distributed}. We trained the \texttt{doc2vec} algorithm using the titles of the posts associated to each considered URL in our dataset. Then, we used the soft K-means clustering algorithm~\cite{kim2007soft} to group the obtained \texttt{doc2vec} vectors into $K$ clusters. We then defined the topic distribution $\gamma_i$ as the soft assignment of item $i$ to the $K$ clusters. We found meaningful results with this method for $K=5$. A textual representation of these topics is shown in Table~\ref{table:reddit-topics}. %
We observed that the identified topics are well differentiated, allowing  to provide intuitive naming as follows: economy, Emailgate, foreign policy, campaigning, and minorities' rights. These five topics are evenly distributed in the data.

\begin{table}[t!]
  \caption{Topics discovered in the Reddit dataset with the \texttt{doc2vec} clustering method, with titles and most common words in each topic. \label{table:reddit-topics}}
  \vspace{-3mm}
  \footnotesize
 \begin{adjustbox}{width=0.9\linewidth}
\begin{tabular}{ll}
\toprule
        \textbf{\textsf{Economy}} &   World, America, Tax, Americans, Income, American \\
      \textbf{\textsf{Emailgate}} &  Emails, Dnc, Russian, Campaign, Email, Foundation \\
 \textbf{\textsf{Foreign policy}} &          Nsa, War, Syria, Government, Russian, Cia \\
       \textbf{\textsf{Campaign}} &      Democrats, Vote, Party, Democratic, Why, Poll \\
     \textbf{\textsf{Minorities}} &            Police, Muslim, Man, Black, Year, Video \\
\bottomrule
\end{tabular}
\end{adjustbox}

  \vspace{-4mm}
\end{table}

In order to evaluate how explainable are the predictions made by our algorithm, and therefore how interpretable are the embeddings we find, we investigate
how well the polarities  $\phi$ estimated from
the content that the nodes' propagate reflect known political leanings of each subreddit\footnote{To reach
  the best possible representation, we used the full reddit dataset,
  and we re-run our model $10$ times and picked the outcome that
  obtained the largest likelihood.}. Given that the topics we
identified (Table~\ref{table:reddit-topics}) fairly represent
different axes of the U.S. political debate and divisions, we expect
these divisions to be reflected in the estimated embeddings.  To do
so, we focus in this analysis on the most active $15$ subreddits, reflecting
discussions to support a U.S. politician or political party.
The political positions typically expressed in these subreddit are well-known and have been analyzed in the literature~\cite{massachs2020roots, soliman2019characterization, roozenbeek2017read, hendricks2017social}.
These include, among the others:

\begin{itemize}
  \item \texttt{democrats}, subreddit affiliated to the
    U.S. Democratic Party. %
  \item \texttt{the\_donald}, a subreddit (now banned) dedicated to supporters of Republican president Donald Trump. %
  \item \texttt{sandersforpresident} and \texttt{wayofthebern} are
    both communities of supporters of Bernie Sanders, democratic
    candidate in 2016 and 2020 primary elections. While the former
    subreddit is official, the second one was created by
    supporters after Sanders lost the primary election in 2016, in
    opposition to Hillary Clinton and the Democratic Party
    establishment.
  \item \texttt{libertarian} is a community focused on libertarianism,
    close but not affiliated to the U.S. Libertarian Party.
\end{itemize}

\newcommand{\lgdcircle}[1]{ \tikz\draw[#1,fill=#1] (0,0) circle (.5ex); }
\definecolor{ourmodel}{HTML}{1f77b4}
\definecolor{barberamodel}{HTML}{ff7f0e}
\definecolor{originalinftopics}{HTML}{2ca02c}
\definecolor{originalinformation}{HTML}{d62728}
\definecolor{node2vectopicsd11}{HTML}{9467bd}
\definecolor{node2vectopicsd128}{HTML}{8c564b}
\definecolor{node2vecd11}{HTML}{e377c2}
\definecolor{node2vecd128}{HTML}{7f7f7f}

\begin{figure*}[t]
  \raisebox{-.5\height}{\includegraphics[width=0.38\linewidth]{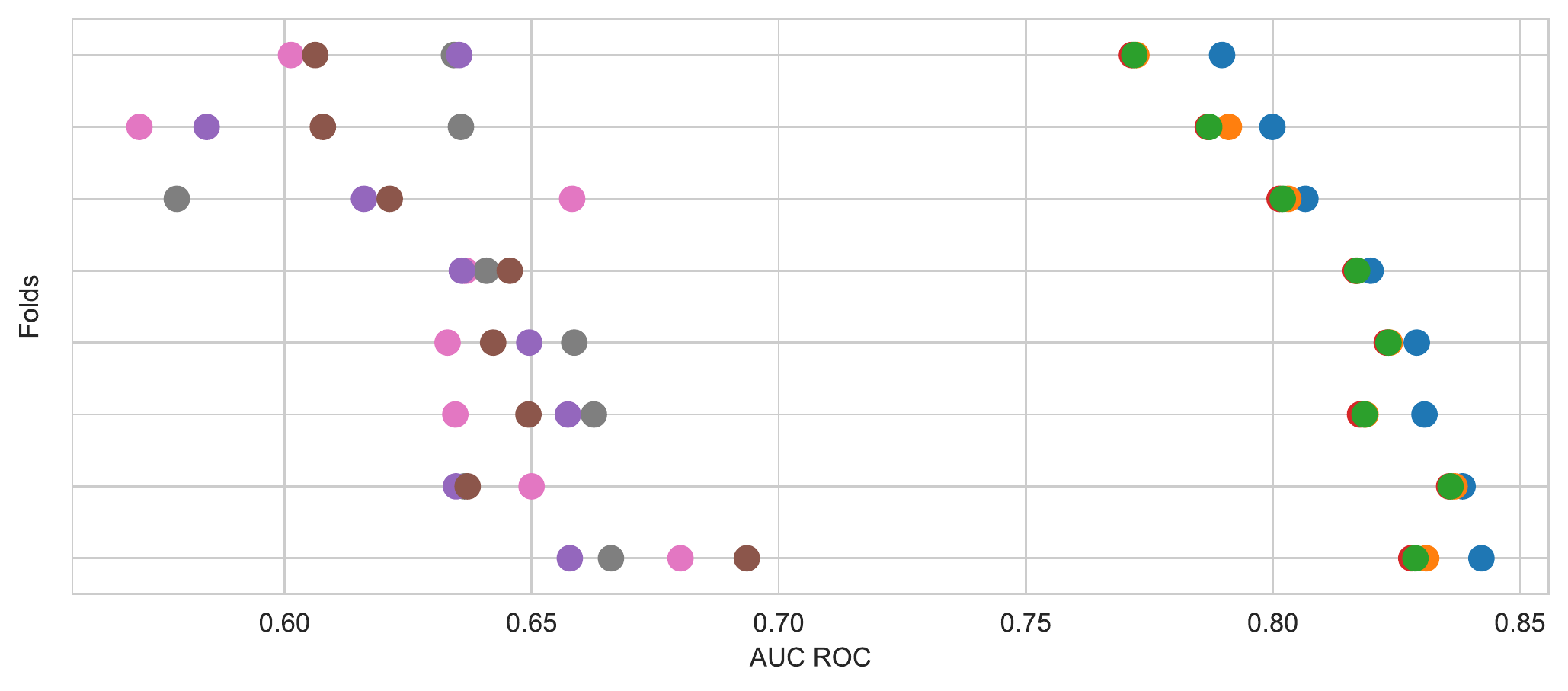}}
  $\quad$
  \fontsize{6.5pt}{7.5pt}\selectfont
  \begin{tabular}{lllllr}
\toprule
                                 &                 Algorithm &            AUC ROC &          Difference &     Avg. Precision &  Time (s) \\
\midrule
            \lgdcircle{ourmodel} &          Our \model model &  $0.820 \pm 0.019$ &   $0.000 \pm 0.000$ &  $0.777 \pm 0.022$ &       7.1 \\
        \lgdcircle{barberamodel} &             Barbera model &  $0.812 \pm 0.022$ &  $-0.008 \pm 0.005$ &  $0.790 \pm 0.018$ &       6.4 \\
   \lgdcircle{originalinftopics} &    Original inf. + Topics &  $0.811 \pm 0.022$ &  $-0.009 \pm 0.006$ &  $0.781 \pm 0.018$ &      10.1 \\
 \lgdcircle{originalinformation} &      Original information &  $0.810 \pm 0.022$ &  $-0.009 \pm 0.006$ &  $0.785 \pm 0.018$ &       9.5 \\
        \lgdcircle{node2vecd128} &           node2vec, d=128 &  $0.639 \pm 0.028$ &  $-0.180 \pm 0.024$ &  $0.660 \pm 0.033$ &      11.6 \\
  \lgdcircle{node2vectopicsd128} &  node2vec + Topics, d=128 &  $0.638 \pm 0.028$ &  $-0.182 \pm 0.016$ &  $0.650 \pm 0.031$ &      14.5 \\
   \lgdcircle{node2vectopicsd11} &   node2vec + Topics, d=11 &  $0.634 \pm 0.024$ &  $-0.186 \pm 0.019$ &  $0.634 \pm 0.033$ &       8.7 \\
         \lgdcircle{node2vecd11} &            node2vec, d=11 &  $0.633 \pm 0.034$ &  $-0.186 \pm 0.024$ &  $0.641 \pm 0.036$ &       7.3 \\
\bottomrule
\end{tabular}

  \normalsize
    \vspace{-2mm}
    \caption{
    Comparison of prediction accuracy, in cross validation, on the Reddit dataset.
    For each metric, we report mean and standard deviation across all folds.
    The \emph{Difference} column reports the difference in AUC ROC between each baseline and our model.
    In the plot on the left, each horizontal line represents a fold; on the X axis, we report the AUC ROC for each model.
    \label{fig:reddit-predictions}}
\end{figure*}

\spara{Results.}
We first consider the topics economy, minorities rights, and Emailgate due to the
ease in their interpretability and their similarity with axis identified as divisive in the 2016 elections (i.e. economy, identity issues, and anti-establishment disaffection~\cite{drutman2017political}).
We report the estimated polarities for
each of these topics in both the plots and the tables of
Figure~\ref{fig:reddit-polarity-plot}.

On the ``economy'' ideological axis, we see that
the most extreme positions among the aforementioned subreddits are
filled by \texttt{libertarian} on one side and
Bernie Sanders' on the other. This accurately represent the deep
division between the pro-business libertarian ideology and the more
pro-welfare ideas of democrats~\cite{drutman2017political}, and even more with the socialist ideas of Bernie Sanders supporters~\cite{mills2018pop}. \texttt{the\_donald} correctly appears to be on
the same side of the axis as libertarians. On the
minorities' rights ideological axis, instead, we found that
\texttt{the\_donald} lies on one extreme, \texttt{democrats} on the
other, while \texttt{libertarian} have more moderate views.
These reflects common readings of the positions of these groups on the economical axis~\cite{drutman2017political}.
On the ``Emailgate'' axis, the divisions reflect how different
communities perceived the controversy: in particular,
\texttt{the\_donald} (together with \texttt{conspiracy} and \texttt{hillaryforprison}) lies on the
very extreme end of the spectrum, while \texttt{democrats} on the
other. \texttt{libertarians} on this axis are very separated from \texttt{the\_donald}. It is known that Donald Trump has been using extensively the so-called
Emailgate controversy to attack Hillary Clinton, while the Democratic
Party defended her; Libertarians
did not attack Clinton on the
topic\footnote{\url{https://www.washingtonpost.com/news/post-politics/wp/2016/07/07/libertarian-rivals-wont-attack-clinton-over-emails/}}.
These positions are all well-represented in our ideological embeddings.
Moreover, we find a separation between \texttt{wayofthebern}, created in opposition with Hillary Clinton's victory of the 2016 primaries, and \texttt{democrats}: the former lies in the middle on the issue, while the latter is in the most extreme spot.
Their embeddings in this axis, again, echo known facts from the literature that studied these communities: \texttt{the\_donald} and \texttt{conspiracy} pushed these accusations, while \texttt{wayofthebern} occasionally ``[tried] to leverage the dual valence of the \emph{deep state} frame -- appealing to the older, more general anti-national-security-establishment frame -- even as the frame continues to do work on the right''~\cite{benkler2018network}.
Moreover, these findings reflect survey measurements on anti-establishment disaffection between Clinton and Sanders supporters~\cite{drutman2017political}.
Finally, beside these three axes, we find that the foreign policy axis is highly correlated with the minorities' rights one.
This matches with previous observations from political surveys~\cite{baumann2020emergence}.

We summarize some of these examples in Figure~\ref{fig:radar-plot}; there, we visualize the position of a subreddit on an axis by indicating its rank among the considered subreddits on that axis.
From these examples, this evaluation shows how our representation of nodes in multidimensional ideological space can provide an intuitive and interpretable representation of the opinion and beliefs of each node.

\subsection{Prediction accuracy on real-world data}
\label{sec:real-world}

\begin{figure*}[t]
  \raisebox{-.5\height}{\includegraphics[width=0.38\linewidth]{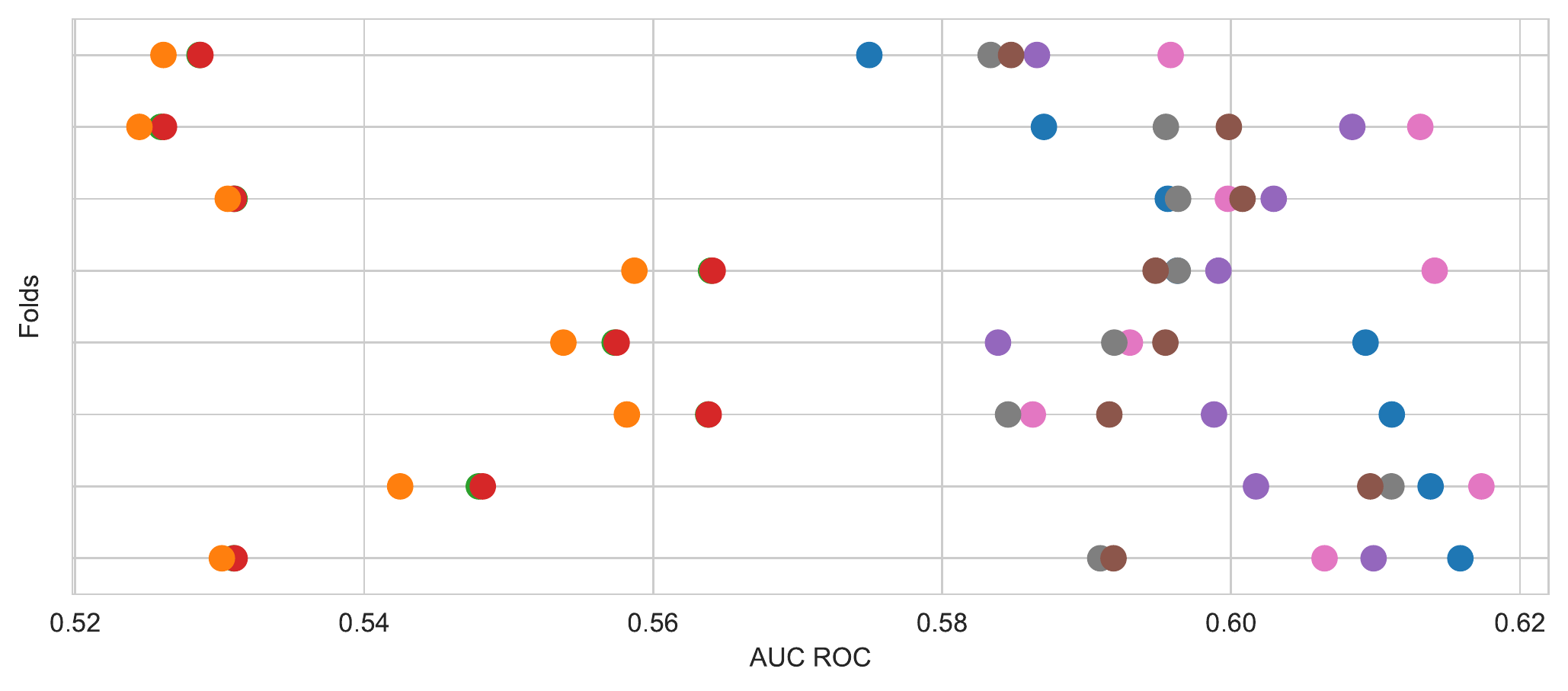}}
  $\quad$
  \fontsize{6.5pt}{7.5pt}\selectfont
  \begin{tabular}{lllllr}
\toprule
                                 &                 Algorithm &            AUC ROC &          Difference &     Avg. Precision &  Time (s) \\
\midrule
            \lgdcircle{ourmodel} &          Our \model model &  $0.601 \pm 0.015$ &   $0.000 \pm 0.000$ &  $0.435 \pm 0.013$ &      83.1 \\
         \lgdcircle{node2vecd11} &            node2vec, d=11 &  $0.603 \pm 0.011$ &   $0.003 \pm 0.018$ &  $0.431 \pm 0.010$ &     480.5 \\
  \lgdcircle{node2vectopicsd128} &  node2vec + Topics, d=128 &  $0.596 \pm 0.007$ &  $-0.004 \pm 0.014$ &  $0.427 \pm 0.008$ &     561.7 \\
   \lgdcircle{node2vectopicsd11} &   node2vec + Topics, d=11 &  $0.599 \pm 0.009$ &  $-0.002 \pm 0.015$ &  $0.425 \pm 0.009$ &     485.3 \\
        \lgdcircle{node2vecd128} &           node2vec, d=128 &  $0.594 \pm 0.009$ &  $-0.007 \pm 0.014$ &  $0.425 \pm 0.006$ &     523.1 \\
   \lgdcircle{originalinftopics} &    Original inf. + Topics &  $0.544 \pm 0.016$ &  $-0.057 \pm 0.016$ &  $0.391 \pm 0.020$ &     985.7 \\
 \lgdcircle{originalinformation} &      Original information &  $0.544 \pm 0.016$ &  $-0.057 \pm 0.016$ &  $0.391 \pm 0.020$ &     965.4 \\
        \lgdcircle{barberamodel} &             Barbera model &  $0.541 \pm 0.015$ &  $-0.060 \pm 0.015$ &  $0.387 \pm 0.019$ &      75.9 \\
\bottomrule
\end{tabular}

  \normalsize
    \vspace{-2mm}
    \caption{Comparison of prediction accuracy, in cross validation, on the Twitter dataset.
    For each metric, we report mean and standard deviation across all folds.
    The \emph{Difference} column reports the difference in AUC ROC between each baseline and our model.
    In the plot on the left, each horizontal line represents a fold; on the X axis, we report the AUC ROC for each model.
    \label{fig:twitter-predictions}}
\end{figure*}

Our final goal is to check whether our inference algorithm is able to capture the real-world activation dynamics and to make accurate predictions on unseen cascades.
In order to evaluate the tradeoff between explainability and predictive power, we compare our model with the following representative popular baselines:%
\begin{description}
  \item[node2vec~\cite{grover2016node2vec}:] we define a weighted graph by considering an arc $(v, u)$ for each instance of ``node $u$ activates on an item $i$ after node $v$''. Then, we embed this graph using the node2vec algorithm into a representation of the same dimension as our ideological space, i.e. $2K$ dimensions (corresponding to $\theta$ and $\phi$). The idea is to check whether the resulting embedding is capable of summarizing the same information provided by our embeddings. %
  We also perform the same experiments with embedding dimension $d=128$, usually employed in literature~\cite{goyal2018graph}.
  After computing these embeddings, we train a logistic regression classifier on the Hadamard product of the embeddings, as suggested by~\citet{grover2016node2vec}.
  We remark that, despite node2vec being one of the most popular and well-performing graph embeddings~\cite{goyal2018graph}, the results it provides are not directly interpretable~\cite{vskrlj2020tax2vec}.
  \item[Barbera's model~\cite{barbera2015birds}:] the goal of this algorithm is to embed nodes of a given graph into a one-dimensional bipolar ideological spectrum (e.g. left-right) which models homophily and divergence. We apply it
    on the same activation graph defined for node2vec.
    This model is usually applied to networks, and it is not specific to information cascades.
    Also, it provides weaker representation capabilities, since its output falls in the mono-axis opinion mining line of research discussed in Section~\ref{sec:related}.
  \item[Original information:] for this approach, we do not use any embedding. We train a logistic classifier directly on a one-hot encoding
    representation of $u$ and $v$ (i.e., in $2 \cdot |\nodes|$ dimensions, a much larger space than the ones we use for embeddings). The classifier is given the concatenated one-hot representations of a pair of nodes and it is trained to predict whether that pair of nodes correspond to a successful propagation.
    The approach is therefore not scalable for very large graphs, and we use it as a naive baseline.
\end{description}
Since our model leverages the topic distribution of each item as
input, we also tested baselines using the same information, by
concatenating the topic distribution of each item to each input vector
of the logistic regression models (node2vec, original
information). Barbera's model, instead, does not offer an immediate
way to use this kind of information.

To train and evaluate each algorithm, we divided the dataset in 10 folds by splitting independent items into equally sized groups.
We used the first two folds as a validation set to tune the parameters of node2vec, $p$ and $q$.
We chose the best parameters according to the average AUC ROC in these two folds.
Then, we used the remaining $8$ folds in cross-validation to assess results.
The evaluation is performed on two datasets, namely the Reddit dataset previously described, and a Twitter dataset we describe in detail next.

\spara{Prediction results on Reddit.} \label{sec:real-world-subsubsec}
Figure~\ref{fig:reddit-predictions} reports a comparison of the results obtained on the Reddit dataset.
On this dataset, our approach results substantially in line with state-of-the-art. Moreover, on each single fold it consistently outperforms the baselines~(\emph{Difference} column in Figure~\ref{fig:reddit-predictions}). In practice, the richness and interpretability of our embeddings, illustrated in the previous section, do not come at the expense of predictive power. The explicit modeling of polarities on ideological axes still enables, in fact, useful predictions of unobserved behavior.
Notably, on this network node2vec falls significantly below all the other baselines.

\spara{Prediction results on Twitter.}
We then proceed to evaluating our predictive performance on a larger dataset representing Twitter resharing behavior in a polarized setting.
For the construction of this dataset, we considered the accounts of
$7152$ Twitter users extracted by~\citet{vaccines},
along with all their posted tweets. These users have mainly posted
about Italian politics hence are in the same weakly connected
component of a retweet network.
We consider as potential propagations all the instances of user $u$ retweeting a tweet $t$ after user $v$.
We manually selected $6$ hashtags
related to public political opinions to be used as topics: three
politicians (\#Salvini, \#DiMaio, \#Renzi) and three debates (\#Vaccini,
\#Migranti, \#Tav). We constructed the dataset from the tweets that
contain these hashtags. From this subset, we iteratively removed users and tweets
with fewer cascades,
so that every Twitter user in the dataset has at least 100 retweets
and every retweet has been shared by at least 10 users. This way, we
obtained the final dataset with $738$ Twitter users and $3624$
retweets.

We define the topic distribution of each item simply by counting the
appearances of hashtags as follows. Denoting $\{ h_1, \dots, h_K \}$
as the set of selected hashtags and $H_i$ as the set of selected
hashtags mentioned by tweet $i$, we define the topic distribution of
tweet $i$ as
$
  \gamma_{i,k} = %
  \frac{1}{ |H_i| } \text{ if } h_k \in H_i
  $ and $0$ otherwise.

The retweet graph, built according to the procedure illustrated in the previous section and needed by node2vec, exhibits an extremely high ($0.91$) density. In order to make it tractable by node2vec, we have to filter out all edges representing a number of retweets less than the average ($5.39$), thus obtaining a final
density of $0.38$.

Figure~\ref{fig:twitter-predictions} reports the results of the evaluation. This dataset seems noisy and
difficult to predict, but still our model
is able to recover some information: we measure an average AUC ROC ($0.601$) significantly higher than the simpler baselines ``original information'' and Barbera's model.
We conjecture that considering different ideological axes in a meaningful way is essential in predicting behavior in such a dataset, confirming that it respects the general assumptions of our model.
In sharp contrast with the previous dataset, node2vec is instead the best performing model, essentially tied to our approach (their AUC ROCs are within one standard deviation from each other).
We highlight, however, the clear advantage of our model in terms of computing time, beside its interpretability.

These two datasets show that our model is able to combine the interpretability of simpler models on smaller datasets with the prediction accuracy of state-of-the-art embeddings on larger ones.

\section{Conclusions and Future Work}
\label{sec:conclusions}

In this paper, taking an information-propagation standpoint, we introduce \emph{ideological embeddings}, 
i.e., the mapping of the ideological leanings of social media users in a multidimensional ideological space.
To this end, we propose a stochastic propagation model, which formalizes how politically salient content spreads in a social network as a function of the ideological embeddings of users under homophily-driven interactions, and we devise a scalable gradient-based optimization procedure to learn the ideological embeddings that maximize an approximation of the likelihood of a set of information cascades.

We show through experiments on real-world as well as synthetic datasets, that the learnt ideological embeddings are coherent with our model, that they are interpretable, and that they can offer reliable and explainable predictions of unseen behavior.

In general, our framework can aid studies on how polarizing content spreads on social networks: since controversy and confirmation bias are important elements of misinformation spread, our model could help to get a better understanding of the topics that are prone to misinformation.
However, \emph{we do not claim that our model is in general the best way to predict unobserved behavior in information cascades}: its performance relies on the validity of its assumptions to a particular context.
Instead, it provides a novel angle to the problem, able to turn raw information cascades into an interpretable embedding with user-defined axes.

Under this perspective it could be extended to adapt to other contexts, or to test different assumptions.
For instance, modeling replies on Twitter instead of retweets would need a different set of assumptions because of their different characteristics in terms of homophily~\cite{lai2019stance}.
Since replies can be antagonizing, propagation can be driven by either alignment or misalignment. Another direction is
to extend our framework to jointly infer the ideological leanings of social media users and content (by directly modeling items' polarity as latent variables), as well as the axes of the ideological space and the underlying topic distribution.

\section*{Acknowledgments}
CM and FB acknowledge support from Intesa Sanpaolo Innovation
Center. GM acknowledges support from the EU H2020 ICT48 project "HumanE-AI-Net" under contract \#952026. 
The funders had no role in study design, data collection
and analysis, decision to publish, or preparation of the manuscript.

\balance

\bibliographystyle{ACM-Reference-Format} %
\bibliography{abbr,references}

\end{document}